\documentclass[10pt,journal]{IEEEtran}
\usepackage{tabularx}
\usepackage[T1]{fontenc}
\usepackage[noend]{algpseudocode}
\usepackage{graphicx}
\usepackage{amsmath,amssymb,amsfonts,stfloats}
\usepackage{array}
\usepackage{textcomp}
\usepackage[font=small]{caption}
\usepackage{paralist}
\usepackage{lettrine}
\usepackage{cite}
\usepackage{xcolor}
\usepackage{cancel}
\usepackage{booktabs}
\usepackage[export]{adjustbox}
\usepackage[caption=false,font=footnotesize]{subfig}
\usepackage{url}

\usepackage{breakurl}
\usepackage{multirow}
\usepackage{multicol}
\usepackage[ruled,vlined]{algorithm2e}
\newcommand{\norm}[1]{\left\lVert#1\right\rVert}
\newcolumntype{C}[1]{>{\centering\arraybackslash}p{#1}}

\begin{document}

\title{Deep Learning-based Target-To-User Association in Integrated Sensing and Communication Systems}
\author{Lorenzo~Cazzella,~\IEEEmembership{Graduate Student Member,~IEEE,} 
Marouan~Mizmizi,~\IEEEmembership{Member,~IEEE,}
Dario~Tagliaferri,~\IEEEmembership{Member,~IEEE,} Damiano~Badini,~\IEEEmembership{Member,~IEEE,}
Matteo~Matteucci,~\IEEEmembership{Member,~IEEE,} \\
and~Umberto~Spagnolini,~\IEEEmembership{Senior Member,~IEEE}
}

\markboth{}%
{}

\maketitle

\begin{abstract}
In Integrated Sensing and Communication (ISAC) systems, matching the radar targets with communication user equipments (UEs) is functional to several communication tasks, such as proactive handover and beam prediction. In this paper, we consider a radar-assisted communication system where a base station (BS) is equipped with a multiple-input-multiple-output (MIMO) radar that has a double aim: \textit{(i)} associate vehicular radar targets to vehicular equipments (VEs) in the communication beamspace and \textit{(ii)} predict the beamforming vector for each VE from radar data. The proposed target-to-user (T2U) association consists of two stages. First, vehicular radar targets are detected from range-angle images, and, for each, a beamforming vector is estimated. Then, the inferred per-target beamforming vectors are matched with the ones utilized at the BS for communication to perform target-to-user (T2U) association. Joint multi-target detection and beam inference is obtained by modifying the you only look once (YOLO) model, which is trained over simulated range-angle radar images. Simulation results over different urban vehicular mobility scenarios show that the proposed T2U method provides a probability of correct association that increases with the size of the BS antenna array, highlighting the respective increase of the separability of the VEs in the beamspace. Moreover, we show that the modified YOLO architecture can effectively perform both beam prediction and radar target detection, with similar performance in mean average precision on the latter over different antenna array sizes.

\end{abstract}


\IEEEpeerreviewmaketitle

\section{Introduction}

\IEEEPARstart{I}NTEGRATED sensing and communication (ISAC) are poised to be a fundamental technology for the forthcoming 6G era \cite{liu2022integrated}. This revolutionary concept combines advanced wireless communication capabilities with pervasive sensing, paving the way for unprecedented levels of connectivity and intelligence in future networks \cite{huang2021mimo, demirhan2022radar, gonzalez2016radar}. Moreover, exploiting higher frequency bands in 6G, e.g., millimeter waves (mmWaves), enables finer sensing resolutions, expanding ISAC's potential. This amalgamation propels advancements in domains such as autonomous systems, smart cities, and healthcare, revolutionizing the interaction with technology. The fusion of communication and sensing transcends mere opportunity, becoming an imperative for harnessing the full potential of 6G networks. 

The design of ISAC technology consists of trading between sensing and communication performance, leaning towards one or the other depending on the target application. The complexity of the tradeoff depends on the level of integration between the two functionalities \cite{zhang2021overview}, stemming from colocation, where sensing and communication shares partially the hardware, to co-design, where they share the same space-time-hardware resources. The work herein can be framed within the communication-centric ISAC framework \cite{liu2022integrated}, where we explore sensing as a secondary function to support communication for beam management in a multi-vehicular scenario. 

Beam management poses significant challenges that hinder the deployment of 5G new radio (NR) at mmWave frequencies \cite{9502647}, which becomes even more complex when considering Vehicle-to-Everything (V2X) applications \cite{mizmizi2022fastening}. Efficient beam management is critical for maintaining reliable and stable connections in dynamic environments. Factors such as link blockages \cite{9838711} and multi-path propagation create hurdles for beam alignment, tracking, and handover \cite{ganji2022beamsurfer}. 

Beam management solutions encompass three main classification methods: blind techniques (e.g., \cite{mizmizi2022fastening, morandi2021probabilistic}), side information-based methods (e.g., \cite{alrabeiah2020deep, khan2020position}), and ISAC approaches (e.g., \cite{demirhan2022radar, gonzalez2016radar}). Blind solutions use pre-designed beam training sequences and exploit the statistical properties of the channel to estimate the optimal beamforming direction iteratively. However, blind techniques introduce significant overhead and can be inefficient. Side information techniques utilize the transmitter and receiver location information for optimal beamforming direction estimation. While these methods can be effective, they require a pre-established link for information sharing.

ISAC approaches have emerged as a promising alternative that attracted significant attention recently. In the context of beam management, the authors in \cite{gonzalez2016radar} introduced radar-aided beam alignment in mmWave V2I communications exploiting side information gathered from a radar mounted at the infrastructure to select the beams of the communication system.
A deep learning-based beam prediction solution has been recently introduced in \cite{demirhan2022radar}, which studies the impact of different preprocessed raw radar measurements---i.e., range-angle maps, range-velocity maps, and radar cube---on the network predictive capabilities. The proposed method is evaluated on real-world mmWave radar and communication data acquired in vehicular environments, showing high accuracy, significant reduction in beam training overhead, and leading performance when using range-angle radar maps as conditioning inputs.

To exploit radar sensing for beam management, two key steps must be addressed: target detection (TD) and target-to-user (T2U) association. TD involves distinguishing the relevant targets from the clutter in the received radar image. On the other hand, the T2U association focuses on determining the correspondence between the detected targets in the radar images and the communication users. The association step is fundamental in establishing the appropriate beamforming direction for each user based on the radar information. By successfully addressing both TD and T2U association, radar sensing can effectively contribute to efficient beam management in wireless communication systems.

TD is a central and challenging problem in computer vision, playing a fundamental role in several applications. A remarkable breakthrough in TD performance has emerged with the proposal of region-based convolutional neural network (R-CNN) architecture, which extracts hierarchical features from CNNs and utilizes region proposals for TD \cite{girshick2014rich}. R-CNN is a \textit{two-stage detector}, which (i) proposes a set of object bounding boxes, and (ii) predicts for each proposed bounding box the presence of an object and its class. By contrast, \textit{one-stage detectors}, e.g., Single Shot Detector (SSD) \cite{liu2016ssd} and YOLO \cite{redmon2016you}, complete the TD task in a single step. Fundamental milestones in two-stage deep learning-based TD are represented by the introduction of spatial pyramid pooling networks (SPPNet) \cite{he2015spatial}, Fast R-CNN \cite{girshick2015fast}, Faster R-CNN \cite{ren2015faster}, and Feature Pyramid Networks (FPNs) \cite{lin2017feature}.

The You Only Look Once (YOLO) architecture has been proposed in \cite{redmon2016you} as the first single-stage object detection model. Renowned for its efficiency, the YOLO model infers bounding boxes in parallel for separate image regions. Nevertheless, it experiences lower localization accuracy compared to the two-stage detectors. Several YOLO versions have been developed to overcome this limitation, leading to the recently proposed YOLOv8 \cite{yolov8}, which represents the state-of-the-art model for both efficiency and accuracy. We refer the interested reader to \cite{zou2023object} for an exhaustive survey on TD models.

YOLO-based multi-target detection algorithms have been successfully applied to radar sensing in the automotive field. A method for simultaneous detection and classification of radar targets in automotive scenarios based on YOLOv3 is proposed in \cite{kim2020yolo}. In \cite{long2020lira}, the authors design a lightweight CNN leveraging dense connections, residual connections, and group convolution and compose it with the YOLO architecture to build a lightweight model for the detection of marine ship objects in radar images, showing good mean average precision (mAP) over two experimentally acquired datasets. A multi-data source YOLO-based TD is in \cite{song2022ms}, where the authors fuse the information derived from a mmWave radar and a camera to improve the detection performance.

The T2U association is a key operation in ISAC systems, but research in this area remains limited. In realistic scenarios, T2U association poses significant challenges due to: \textit{i)} the presence of multiple users and targets, potentially resulting in a non-empty set difference; \textit{ii)} differing resolutions in sensing and communication information---for example, sensing provides angle, range, and Doppler, while communication yields beam and received power data; \textit{iii)} practical considerations, where vehicles, treated as extended targets, undergo geometric distortions known as foreshortening and overlay \cite{mizmizi2023target} when illuminated by a sensing apparatus positioned at heights of base station (BS). These distortions introduce biases in target position estimation, leading to mismatches between communication and sensing information.
The most notable works addressing this topic are \cite{aydogdu2020distributed, wang2022multi, mizmizi2023target}. In \cite{aydogdu2020distributed}, the focus is on sensing-aided vehicular communications, where the challenge lies in associating vehicle ID with detected targets using radar and GPS data. The authors utilize the Kullback-Leibler divergence as a similarity metric to solve the constrained data association problem. Nevertheless, the accuracy and reliability of the proposed approach can be affected by environmental factors and by the need for an active low-rate uplink channel for GPS data exchange.
On the other hand, \cite{wang2022multi} proposes a different approach that overcomes the aforementioned limitations. They utilize ISAC-only data for multi-vehicle tracking and perform ID association based on the Kullback-Leibler divergence between estimated and predicted vehicle states (range, velocity, and angle).
In a more recent work \cite{mizmizi2023target}, the authors introduce reconfigurable intelligent surfaces (RIS) mounted on vehicles' roofs and strategically configured to reflect the sensing signal towards the radar with a known pattern, enabling accurate T2U association and improving TD. 

The abovementioned works require additional hardware \cite{mizmizi2023target} or side information \cite{aydogdu2020distributed}. The work in \cite{wang2022multi}, instead, operates on the assumption of successful TD, which is a particularly challenging task for extended targets in highly dynamic scenarios.
Differently from the current state of the art, in the context of ISAC systems, our paper proposes a unified framework for multi-user TD and T2U association to enhance beam management. Indeed, when the association between observed radar target and communication equipment is known at some time steps, beam prediction for a VE can be naturally conditioned on the radar target.

In this paper, we propose the following contributions:
\begin{itemize}
    \item Considering a hybrid MIMO communication system and a mmWave MIMO radar, we employ the latest version of the YOLO architecture (YOLOv8 \cite{yolov8}) to jointly achieve real-time radar multi-target detection and analog beam prediction at the BS for each detected radar target. We model the beam prediction problem as a multi-class classification task over a fixed-size codebook of beamforming vectors, and we integrate the beam classification within the YOLO prediction heads. The resulting model is trained and evaluated over realistic simulated radar range-angle images and accurate ray-tracing wireless channel data. We show that YOLO achieves considerable detection and classification performance on both radar target detection and beam prediction tasks.
    \item Leveraging the beam prediction obtained from YOLOv8, and the beamforming vectors selected for each VE at the BS, we tackle the radar target-to-user (T2U) association problem in the beamspace, showing that the probability of correct association significantly increases with the antenna array size at the BS---which highlights the related increase in the separability of the VEs in the beamspace required for effective association.
    \item We design a framework for the simulation of radar images and wireless communication channels at the communication infrastructure. Dynamic vehicular simulations are achieved integrating the Simulation of Urban Mobility (SUMO) \cite{sumo} vehicular traffic simulator, and the EM simulation software Remcom Wireless InSite \cite{wireless_insite} and Remcom WaveFarer \cite{wavefarer}---used, respectively, for high-fidelity wireless channel and radar channel simulations.
\end{itemize}

The paper is organized as follows: Section \ref{sec:system_model} presents the reference communication system, channel model, and radar signal models; Section \ref{sec:proposed_method} introduces the proposed beamspace Target-to-User association method relying on DL-based radar target detection and beam prediction; Section \ref{sec:results} details the simulation framework and shows the achieved results, while Section \ref{sec:conclusion} reports the conclusions.

\subsection*{Notation} Matrices are denoted by bold upper-case letters, while lower-case letters describe column vectors. $\mathbf{A}^{\mathrm{T}}$, $\mathbf{A}^{*}$, $\mathbf{A}^{\mathrm{H}}$ and $\norm{\mathbf{A}}$ indicate, respectively, the matrix transpose, conjugate, conjugate transpose and Frobenius norm. A Gaussian multi-variate circularly complex random variable $\mathbf{a}$ is denoted With  $\mathbf{a}\sim\mathcal{CN}(\boldsymbol{\mu},\mathbf{C})$ with mean $\boldsymbol{\mu}$ and covariance $\mathbf{C}$. $\mathbb{R}$ and $\mathbb{C}$ represent the sets of real and complex numbers, respectively.

\section{System Model}\label{sec:system_model}

\begin{figure}[t]
    \centering
    \subfloat[][]{\includegraphics[width=.9\columnwidth]{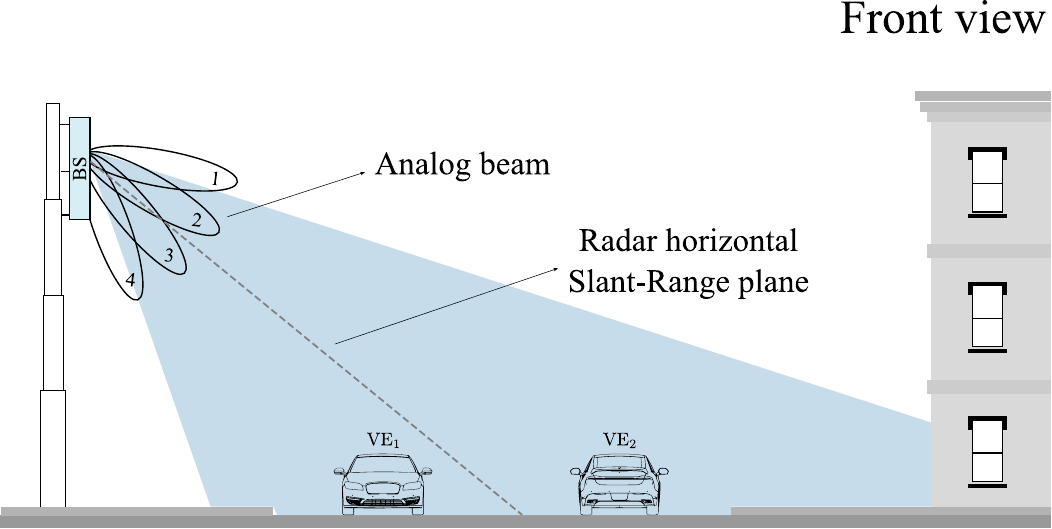}}\\
    \subfloat[][]{\includegraphics[width=.9\columnwidth]{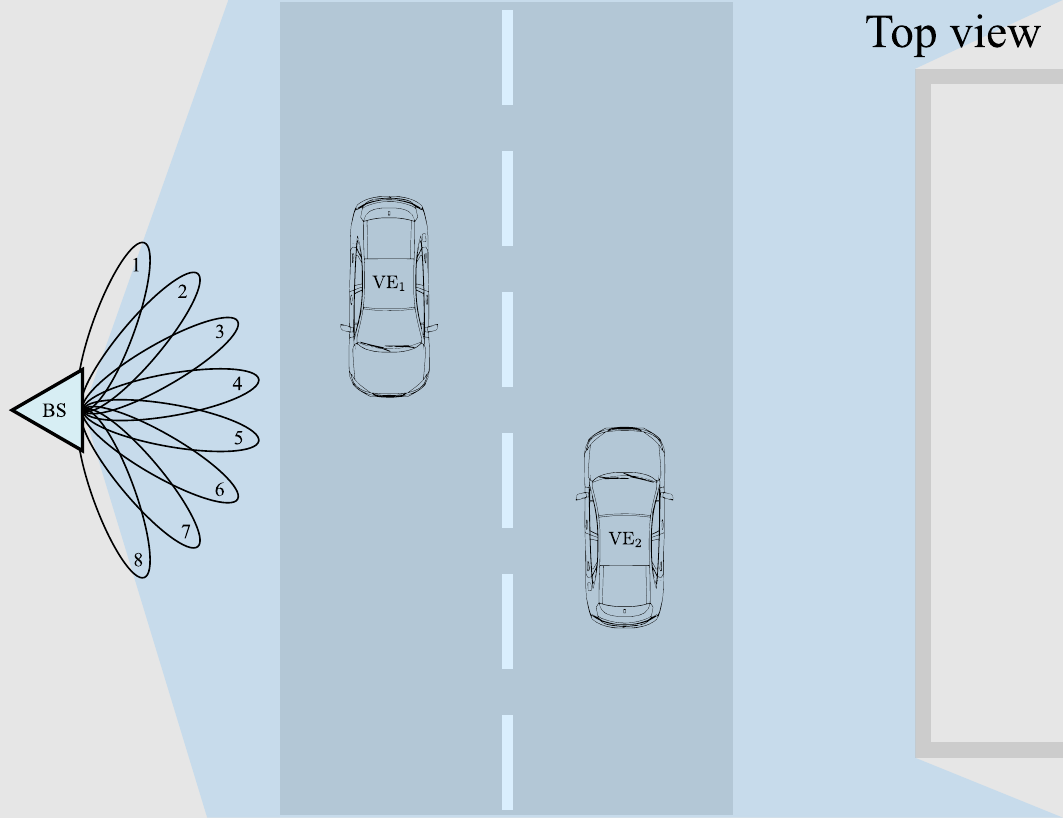}}\\
    \caption{ISAC system with co-located mmWave radar sensor and communication base station. (a) provides a front view representation of the system, showing analog communication beams on the vertical direction at the BS, and highlighting the radar slant-range plane used for radar imaging. (b) shows a top view of the system, depicting analog communication beams on the horizontal direction at the BS.}
    \label{fig:radar_scenario_representations}
\end{figure}

Let us consider the downlink ISAC system depicted in Fig. \ref{fig:radar_scenario_representations}, where a BS is equipped with a mmWave MIMO radar for sensing and a hybrid sub-connected MIMO array for communication. Fig. \ref{fig:radar_scenario_representations} depicts the ISAC system over two different views, highlighting the radar placement and the analog communication beams at the BS. In the following, we detail the model of the communication system and of the radar system.

\subsection{Communication system model}\label{sec:communication_model}

\begin{figure}[t]
    \centering
    \includegraphics[width=\columnwidth]{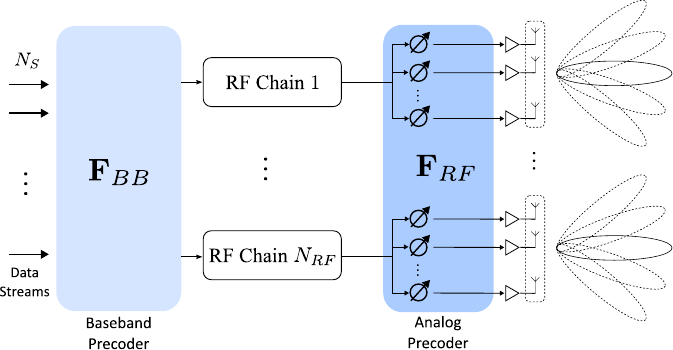}
    \caption{Hybrid MIMO sub-connected Tx. architecture.}
    \label{fig:hybrid_architecture}
\end{figure}
We consider a hybrid mmWave MIMO system where the BS is equipped with a planar antenna array with $N_T = N_T^h \times N_T^v$ antenna elements (along horizontal and vertical directions, respectively) and $N_T^\mathrm{RF}$ RF chains and is using $N_S$ data streams to communicate with $K$ VEs. Each VE is equipped with a planar antenna array with $N_R = N_R^h \times N_R^v$ antenna elements and $N^\mathrm{RF}_R$ RF chains. A sub-connected hybrid MIMO configuration is considered, where the antennas are grouped into sub-arrays of $N_T^B = N_T / N_T^\mathrm{RF}$ and $N_R^B = N_R / N_R^\mathrm{RF}$ antenna elements at the BS and at the VEs, respectively. The transmitted $N_S$ complex-valued symbols are such that $\mathbf{s} \in \mathbb{C}^{N_S\times 1}\sim \mathcal{CN}\left(\mathbf{0}, \mathbf{I}_{N_S}/N_S\right)$, and they are spatially precoded using the cascade of digital and analog precoders $\mathbf{F} = \mathbf{F}_\mathrm{RF} \; \mathbf{F}_\mathrm{BB}$, where $\mathbf{F}_\mathrm{BB}\in \mathbb{C}^{N^\mathrm{RF}_{T} \times N_S}$ is the base-band digital precoder and $\mathbf{F}_\mathrm{RF} \in \mathbb{C}^{N_{T}\times N^\mathrm{RF}_{T}}$ is the RF analog counterpart, obtaining the time-discrete transmitted signal
\begin{equation}\label{eq:Analog&DigitalPrec}
    \mathbf{x} = \mathbf{F}\; \mathbf{s}.
\end{equation}
We assume analog precoding $\mathbf{F}_\mathrm{RF}$ implemented with phased shifters. Hence, its elements are constrained to have the same norm, i.e., $[\mathbf{F}^{(i)}_\mathrm{RF}\mathbf{F}^{(i),\mathrm{H}}_\mathrm{RF}]_{k,k} = 1/N_T$. Moreover, the Tx total power constraint is enforced by designing $\mathbf{F}_\mathrm{BB}$ such that $\norm{\mathbf{F}_\mathrm{RF}\mathbf{F}_\mathrm{BB}}^2 = N_S$ \cite{6717211}. 
 In the sub-connected hybrid configuration, depicted in Fig. \ref{fig:hybrid_architecture}, the analog precoding matrix $\mathbf{F}_\mathrm{RF}$ is block-diagonal
\begin{equation}\label{eq:Frf_subconnected}
    \mathbf{F}_\mathrm{RF} = \begin{bmatrix}\mathbf{f}^{(1)}_\mathrm{RF} & \mathbf{0} &\cdots & \mathbf{0}\\
    \mathbf{0} & \mathbf{f}^{(2)}_\mathrm{RF} & \cdots & \mathbf{0}\\
    \vdots & \vdots & \vdots & \vdots\\
    \mathbf{0} &  \cdots & \mathbf{0} & \mathbf{f}^{(N_T^{RF})}_\mathrm{RF}\end{bmatrix},
\end{equation}
where $\mathbf{0}\in\mathbb{C}^{N_T^B \times 1}$ is a vector with zero-elements and $\mathbf{f}^{(n)}_\mathrm{RF}\in\mathbb{C}^{N_T^B \times 1}$, $n=1,\dots,N_T^{RF}$ denotes the precoding vector for the $n$-th Tx sub-array.

After time-frequency synchronization, the received signal at the $k$th VE can be expressed as
\begin{equation}\label{eq:propagation}
    \mathbf{y}_k = \mathbf{H}_k \mathbf{x} + \mathbf{n}_k,
\end{equation}
where $\mathbf{H}_k \in \mathbb{C}^{N_R \times N_T}$ is the spatially sparse and block-faded MIMO channel between the BS and the $k$th VE, while $\mathbf{n}_k \sim \mathcal{CN}\left(\mathbf{0}, \sigma_n^2 \, \mathbf{I}_{N_R} \delta_{k-u}\right)$ denotes the additive white noise, spatially uncorrelated across the antennas and among different VEs. The channel at mmWave follows the block-fading spatially sparse model \cite{6834753}:
\begin{equation}
    \mathbf{H}_k = \frac{1}{\sqrt{\rho_k}} \sum_{p=1}^{P} \; \alpha_p e^{j2\pi\nu_p t} \; \mathbf{a}_{UE}(\boldsymbol{\vartheta}_p)  \; \mathbf{a}_{BS}^\mathrm{H}(\boldsymbol{\varphi}_p) 
\end{equation}
where $\rho_k$ denotes the average power path-loss of the $k$th UE, $P$ is the number of paths, $\alpha_p \sim \mathcal{CN}(0, \sigma_p^2)$ denotes the Rayleigh distributed complex amplitude of the $p$th path, such that $\sum_p \sigma_p^2 = 1$, $\nu_p$ is the Doppler shift of the $p$th path, $\boldsymbol{a}_{BS}$ and $\boldsymbol{a}_{UE}$ are, respectively, the BS and UE antenna array response vectors, functions of directions of arrival $\boldsymbol{\vartheta}_p$ and directions of departure $\boldsymbol{\varphi}_p$.

At the receiver, the $k$th VE applies the cascade of analog and digital combiners $\mathbf{w}_k = \mathbf{W}^k_\mathrm{RF} \; \mathbf{w}^k_\mathrm{BB}$, with $\mathbf{W}^k_\mathrm{RF} \in \mathbb{C}^{N_{R}\times N^\mathrm{RF}_{R}}$ and $\mathbf{w}^k_\mathrm{BB} \in \mathbb{C}^{N_{R}^\mathrm{RF}\times 1}$. Similarly, the analog combiner $\mathbf{W}_{RF}$ is subject to the same constraint of $\mathbf{F}_{RF}$, i.e., $[\mathbf{W}^{(j)}_{RF}\mathbf{W}^{(j),\mathrm{H}}_{RF}]_{l,l} = 1/N_R$. The received symbol can be expressed as
\begin{equation}\label{eq:Analog&DigitalComb}
\begin{split}
    \hat{s}_k &= \mathbf{w}_k^{\mathrm{H}} \; \mathbf{H}_k \; \mathbf{x} + \mathbf{w}_k^{\mathrm{H}} \; \mathbf{n} \\
    &= \mathbf{w}_k^{\mathrm{H}} \; \mathbf{H}_k \; \mathbf{f}_k \; s_k  + 
    \underbrace{\sum_{i \neq k} \mathbf{w}_k^{\mathrm{H}} \; \mathbf{H}_k \; \mathbf{f}_i \; s_i}_{\mathrm{Intra-Cell\, Interference}} 
    + \underbrace{\mathbf{w}_k^{\mathrm{H}} \; \mathbf{n}}_{\mathrm{Noise}} \\
\end{split}
\end{equation}
where $\mathbf{f}_k \in \mathbb{C}^{N_T \times 1}$ is the cascade of digital and analog precoders of the $k$th UE. Herein, we propose a DL approach in the ISAC context to predict analog precoders $\mathbf{F}_{RF}$. Derivation of $\mathbf{F}_{BB}$, $\mathbf{W}_{BB}, \mathbf{W}_{RF}$ is based on state-of-the-art methods, e.g., \cite{mizmizi2021channel}.

\subsection{Radar signal model and image synthesis}\label{sec:radar_model}

We consider a frequency-modulated continuous wave (FMCW) MIMO radar at the BS side. In the general case, the MIMO radar is equipped with $L_{\mathrm{Tx}}$ Tx antennas and $L_{\mathrm{Rx}}$ Rx ones, for a total of $L=L_{\mathrm{Tx}}L_{\mathrm{Rx}}$ of virtual measurement channels (Tx-Rx pairs). The 3D positions of the Tx and Rx antennas forming the $\ell$-th channel are denoted by $\mathbf{p}_{\mathrm{Tx},\ell}$ and $\mathbf{p}_{\mathrm{Rx},\ell}$, respectively. We assume that the MIMO radar operates in time-division multiplexing, thus each Tx antenna emits a modulated chirp signal, centered around carrier frequency $f_0$, of bandwidth $B_s$ and periodic by the pulse repetition interval (PRI) $T_P$ \cite{Zaugg2015_FMCWSAR}\footnote{The waveforms at the $L_{\mathrm{Tx}}$ Tx antennas are orthogonal in time, but we drop the antenna index and the pulse delay among the antennas for simplicity.}
\begin{align}\label{eq:TXsignal}
x_s(t) = \begin{cases}
      A \, e^{j (2\pi f_{0} t+ \pi \mu t^{2})} & 0 \leq t \leq T_c\\
      0 & T_c \leq t \leq T_P
    \end{cases}     
\end{align}
where $A$ is the amplitude of the Tx signal, $\mu = B_s / (2 T_c)$ denotes the chirp rate, $T_c$ denotes the chirp duration, during which the carrier frequency is linearly increased from $f_0$ up to $f_0 + B_s$. Within the coherent processing interval (CPI)\footnote{The CPI is defined as the time interval within which the Doppler shift of the target (thus its velocity) does not change, thus the Rx echoes can be \textit{coherently} processed.}, the Rx signal for the $\ell$-th measurement channel, pertaining to the $k$-th PRI, after demodulation and mixing with the conjugate chirp $e^{-j (2\pi f_{0} t+ \pi \mu t^{2})}$, is:
\begin{equation}\label{eq:Rx_signal_radar}
\begin{split}
    y^k_\ell(t) & = \sum_{q = 1}^Q \underbrace{\beta^k_{q,\ell} \, e^{-j (2\pi f_{0} \, \tau_{q,\ell} - \pi \mu \tau_{q,\ell}^{2} + 2 \pi \mu \tau_{q,\ell} t  )}}_{s_\ell^k(t; \mathbf{p}_q)} + z^k_\ell(t)
\end{split}
\end{equation}
where \textit{(i)} the number of targets in the environment is denoted by $Q$; \textit{(ii)} $\beta^k_{q,\ell}$ is the complex scattering amplitude the $q$th target, comprising the path-loss, the reflectivity of the target (radar cross section, RCS) and the target's proper scattering phase (i.e., the phase modeling the interaction of the EM wave with the target); \textit{(iii)} $ \tau_{q,\ell}$ is the two-way delay from the Tx antenna to the target and back to the Rx antenna, while \textit{(iv)} $z^k_\ell(t)\sim\mathcal{CN}(0, \sigma_z^2 \delta_{\ell-m})$ is the additive white Gaussian noise, uncorrelated in space (across channels). We denote with $s_\ell^k(t; \mathbf{p}_q)$, for convenience, the Rx signal due to the $q$-th target located in $\mathbf{p}_q$. The delay is 
\begin{equation}\label{eq:delay_radar}
     \tau_{q,\ell} = \frac{\| \mathbf{p}_q-\mathbf{p}_{\mathrm{Tx},\ell}\| + \|\mathbf{p}_{\mathrm{Rx},\ell} - \mathbf{p}_q\|}{c}. 
\end{equation}
The model of the scattering amplitude is \cite{skolnik}
\begin{equation}\label{eq:scattering_amplitude}
        \beta^k_{q,\ell} = \sqrt{\frac{(4 \pi)^{-3}\lambda_0^2 G_\ell(\boldsymbol{\vartheta}_{q,\ell})G_\ell(\boldsymbol{\vartheta}_{q,\ell}) }{ \| \mathbf{p}_q-\mathbf{p}_{\mathrm{Tx},\ell}\|^2 \|\mathbf{p}_{\mathrm{Rx},\ell} - \mathbf{p}_q\|^2 } \Gamma_{q}}\;e^{j (\delta_q + 2\pi \nu_{q} k T_P)} 
\end{equation}
where \textit{(i)} $\lambda_0 = c/f_0$ is the carrier wavelength, \textit{(ii)} $G_\ell(\boldsymbol{\vartheta}_{q,\ell})$ is the power gain of the single radar antenna element in the direction of the $q$th target $\boldsymbol{\vartheta}_{q,\ell}$, \textit{(iii)} $\Gamma_q$ is the radar cross section (RCS) of the target, \textit{(iv)} $\delta_q \sim \mathcal{U}[0,2\pi)$ is a time-invariant phase term within the CPI, modeling both the Tx/Rx circuitry as well as the interaction of the incident EM wave with the $q$th target and \textit{(v)} $2\pi \nu_{q} k T_P$ is the linear phase term with the slow time function of the Doppler shift $\nu_q = 2 V_{q}/\lambda_0$ (for radial velocity $V_q$). As the target's phase $\delta_q$ is not known a-priori, it is customary to model it as an unknown deterministic variable, following a Swerling 0 model \cite{skolnik}.
In \eqref{eq:Rx_signal_radar}, the $q$th target manifests with a time-invariant term comprising the propagation phase $2 \pi f_0 \tau_{q,\ell}$ as well as a quadratic phase term w.r.t. the delay $-\pi \mu \tau_{q,\ell}^2$\footnote{The quadratic phase term $-\pi \mu \tau_{q,\ell}^2$ is sometimes referred to as "video phase" in the radar jargon, and it usually negligible for short-range radars (up to few tens to a hundred of meters in range) compared to the propagation phase.} and a time-varying linear phase term whose frequency is proportional to the delay $\tau_{q,\ell}$.

\textbf{Radar image synthesis}: The Rx data \eqref{eq:Rx_signal_radar} is then processed to generate radar images. A radar image is a 2D or 3D map of the complex reflectivity of the environment, that shall be characterized by the highest possible resolution, i.e., the capability of distinguishing two closely spaced point targets. The image at the $k$ PRI is synthesized by back-projection (BP) on a pre-defined grid of pixels $\{\mathbf{x}\}$. BP consists of a 2D correlation (over time $t$ and channels $\ell=1,...,L$) between the gathered radar data $y^k_{\ell}(t)$ and the theoretical (noiseless) response of a fictitious target located in a given location in space $\mathbf{x}$, i.e., $s_\ell^k(t;\mathbf{x})$: 
\begin{equation}\label{eq:BP}
    I_k(\mathbf{x}) = \sum_\ell \int_{T_P} y^k_{\ell}(t) s_\ell^{k,*}(t;\mathbf{x}) \; dt 
\end{equation}
where $I_k(\mathbf{x})\in \mathbb{C}$ denotes the complex image value at pixel $\mathbf{x}$.
The BP technique for image synthesis is based on the wavefield migration integral extensively used in remote sensing to find the location of one or more targets \cite{Cafforio1991}, and it describes the matched filtering operation along space and time of the environment composed of $Q$ point scatteters.

The image formation procedure \eqref{eq:BP} is split into range compression (RC) and focusing. \textit{Range compression} is the matched filter along time, and, for FMCW radars, it consists of \textit{(i)} a Fourier transform over time \textit{(ii)} a frequency scaling factor $f=\mu t$ \textit{(iii)} a phase rotation to remove the video phase $e^{-j \pi \mu \tau^2}$. 
The RC Rx signal is: 
\begin{equation}\label{eq:Rx_signal_radar_RC}
\begin{split}
    y^k_{\mathrm{RC},\ell}(t) & = \int_{T_P} y^k_\ell(t) e^{-j2 \pi f t}\, dt \bigg \rvert_{f=\mu t} \times e^{j \pi \mu t^2} = \\
    & = \alpha\sum_{q=1}^Q  \beta^k_{q,\ell} \, \mathrm{sinc}\left[B_s (t-\tau_{q,\ell})\right] e^{-j 2 \pi f_0 \tau_{q,\ell}} + n_\ell^k (t)
\end{split}
\end{equation}
where $\alpha = T_c B_s A$,  $\mathrm{sinc}[x]=\sin (\pi x)/(\pi x)$ and $n_\ell^k (t)$ denotes the noise after RC. The result by \eqref{eq:Rx_signal_radar_RC} is a linear combination of scaled ad delayed sinc pulses, multiplied by the propagation phase term $e^{-j 2 \pi f_0 \tau_{q,\ell}}$.
After RC, \textit{focusing} refers to the matched filtering along space, i.e., the available measurement channels. The image $I_k(\mathbf{x})$ is finally formed as:
\begin{equation}\label{eq:image}
\begin{split}
    I_k(\mathbf{x}) & = \sum_\ell y^k_{\mathrm{RC},\ell}(t=\tau_{\ell}(\mathbf{x})) \, e^{+j 2 \pi f_0 \tau_{\ell}(\mathbf{x})} \\
    & \simeq \alpha L \sum_{q=1}^Q  \widetilde{\beta}_{q}^k\,  H_q(\mathbf{x}- \mathbf{p}_q) + W_{k}(\mathbf{x})
\end{split}
\end{equation}
where \textit{(i)} $\widetilde{\beta}_{q}^k$ is the complex amplitude of the $q$-th target in the final image, \textit{(ii)} the two-way delay between $\ell$-th channel and pixel location $\mathbf{x}$ is 
\begin{equation}
     \tau_{\ell}(\mathbf{x}) = \frac{\| \mathbf{x}-\mathbf{p}_{\mathrm{Tx},\ell}\| + \|\mathbf{p}_{\mathrm{Rx},\ell} - \mathbf{x}\|}{c} 
\end{equation}
\textit{(iii)} $H_q(\mathbf{x})$ is the spatial signature of the $q$-th target evaluated over the grid of pixels $\{\mathbf{x}\}$ while \textit{(iv)} $W_{k}(\mathbf{x})$ is the noise in the spatial domain\footnote{$H_q(\mathbf{x})$ is sometimes referred to as spatial impulse response function, meaning the image of a point target (impulse in the spatial domain) obtained with a specific setup, i.e., array aperture and employed bandwidth. For monostatic setups with linear arrays of antennas, $H(\mathbf{x})$ is the mapping in Cartesian coordinates of a 2D sinc in polar coordinates (range and angle).}. BP is a generalization of the conventional DFT-based image synthesis, accounting for any curvature of the wavefront and employed bandwidth, thus it is inherently suited for image synthesis when employing very large arrays, wide bandwidths, and/or non-regular array geometries.  Moreover, differently from DFT-based image synthesis, that provides the image in range-angle(s) domain, BP allows choosing the pixel grid $\{\mathbf{x}\}$ on which to generate the image, accomplishing different imaging possibilities (on both slant planes, i.e., tilted as the radar, and/or ground-level planes). The BP consists of three steps: 1) interpolation of RC sampled data in $t=\tau_{\ell}(\mathbf{x})$, 2) a phase rotation to compensate for the carrier propagation phase by $e^{+j 2 \pi f_0 \tau_{\ell}(\mathbf{x})}$ and 3) the summation (integration) over the available measurement channels. Image \eqref{eq:image}, as any matched filter, can be also cast as an operation that provides the likelihood of the targets' positions in space, from which it follows their estimation by maximum likelihood approaches as described in \cite{5739256}.

\section{YOLO-based Multi Target Detection}\label{sec:YOLO_model}
\begin{figure*}[t!]
    \centering
    \includegraphics[width=\textwidth]{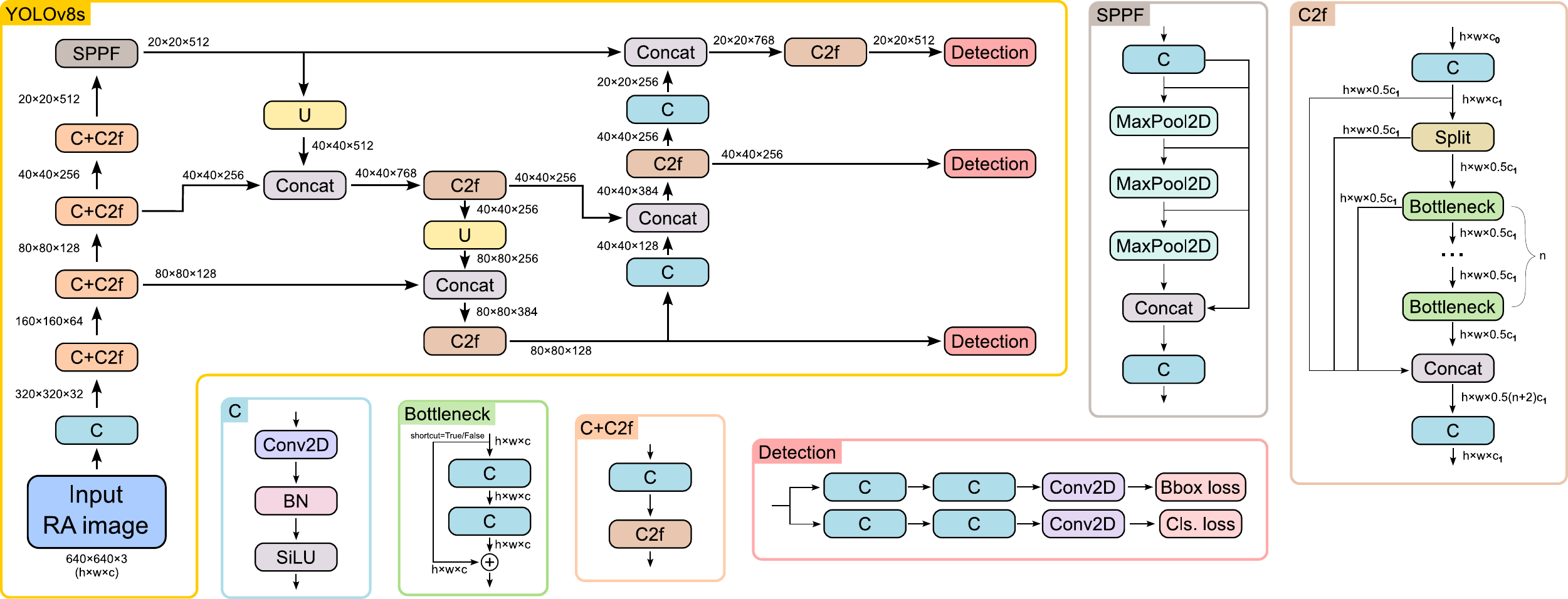}
    \caption{YOLOv8 architecture utilized for target class and beam indices prediction (upper-left block). The remaining blocks expand the layer components depicted in the architecture graph. Conv2D denotes a 2-dimensional convolution, BN indicates batch normalization, U describes an upsampling layer, and SiLU represents a Sigmoid Linear Unit activation function.}
    \label{fig:yolo_architecture}
\end{figure*}

YOLOv8 \cite{yolov8} is a real-time multi-target detection and classification model. It has proven an extremely flexible architecture providing state-of-the-art performance over complex target detection benchmarks. 
The architecture is proposed in multiple model scales that flexibly adapt to heterogeneous application scenarios.
YOLOv8 relies on the successful architectures of the previous YOLO versions and introduces new advancements to improve performance and flexibility \cite{yolov8}. As a result, it leads to state-of-the-art detection and classification accuracy while keeping real-time prediction performance. Its architecture is depicted in Fig. \ref{fig:yolo_architecture} and can be decomposed into backbone, neck, and prediction heads.

\textbf{Backbone}: The objective of the YOLOv8 backbone is to perform feature extraction from the input image. 
It leverages Cross-Stage Partial (CSP) connection blocks \cite{wang2020cspnet} to improve the CNN performance and lower the computational complexity. YOLOv8 introduces the \textit{C2f} module, improving YOLOv5's \textit{C3} module by means of the ELAN design adopted also in \cite{wang2023yolov7}. 
Moreover, YOLOv8 reduces the number of blocks in each stage and the last backbone stage uses the SPPF module, an improved version of SPP \cite{he2015spatial}.

\textbf{Neck:} YOLOv8 performs a multi-scale fusion of features by means of Feature Pyramid Network (FPN) \cite{lin2017feature} and Path Aggregation Network (PAN) \cite{liu2018path} architectures, introduced to counteract the reduction in object location information emerging in \textit{deep} convolutional networks, and the information loss over small objects resulting from the high number of convolution operations. As presented in Fig. \ref{fig:yolo_architecture}, the architecture neck fuses features learned at different scales. 

\textbf{Head:} YOLOv8 is anchor-free, i.e., it directly infers the bounding boxes' centers instead of the offset from the center of a known anchor box. Each prediction head is further split into two separate branches for detection and classification. In Fig. \ref{fig:yolo_architecture}, the two separate prediction heads are shown within the \textit{Detection} sub-module. The YOLO architecture proposes a set of detection bounding boxes and, for each of them, the inferred object classes. During model training, the matching of the model predictions to the ground truth values is performed by means of the Task Aligned Assigner proposed in \cite{feng2021tood}, which evaluates the alignment between inferred information and target labels relying on the weighted scores of bounding box regression and targets classification:
\begin{equation}
    a = p^\lambda \cdot r^\mu,
\end{equation}
where $p$ is the predicted score for the label target class, $r$ is the intersection-over-union (IoU) score  between the predicted bounding box and the ground truth one, and $\lambda$, $\mu$ are tunable hyperparameters.
At inference time, the inferred bounding boxes are post-processed using Non-Maximum Suppression (NMS), aimed at merging bounding boxes that may belong to the same object. NMS is a fundamental step in modern object detection algorithms to improve the detection performance.

The YOLOv8 model is divided into multiple detection branches operating on representations learned from the input image at different scales, to be able to detect objects of different sizes in the image and to allow effective processing of images of different sizes. Each branch is, in turn, split into separate regression and classification heads. The classification heads employ the binary cross-entropy (BCE) loss
\begin{equation}\label{eq:BCE_loss}
    \mathcal{L}_{BCE} = -w(y_n \log x_n + (1 - y_n) \log(1 - x_n)),
\end{equation}
where $w$ is the weight, $y_n$ is the labelled value, and $x_n$ if the value predicted by a YOLOv8 classification head. The BCE loss is commonly used for multi-class classification as it deals with the targets' classes separately and allows to provide during inference a classification score for each possible class

Bounding box regression heads are endowed with Distribution Focal Loss (DFL) and Complete IoU (CIoU) losses. Focal loss accounts for the class imbalance problem common in object detection tasks. Distribution Focal Loss \cite{li2023generalized} is a special case of Generalized Focal Loss, which has been shown to flexibly model the underlying distribution of bounding boxes, leading to more informative and accurate bounding box locations. DFL aims to focus on enlarging the probabilities of the values around target $y$ (i.e., $y_i$ and $y_{i + 1}$). It is defined by
\begin{equation}
    \mathcal{L}_{DFL}(P_i, P_{i + 1}) = -((y_{i+1} - y) \log (P_i) + (y - y_i) \log (P_{i+1})),
\end{equation}
with 
\begin{equation}
    P_i = \frac{y_{i + 1} - y}{y_{i + 1} - y_i};\; P_{i + 1} = \frac{y - y_i}{y_{i + 1} - y_i}.
\end{equation}

We refer the reader to \cite{li2023generalized} for an extensive description of DFL and its derivation from the Generalized Focal Loss continuous framework.

Complete IoU (CIoU) and Distance IoU (DIoU) metrics have been recently introduced in \cite{zheng2020distance} as a possible solution to the slow convergence and inaccurate regression problems of IoU and generalized IoU (GIoU) losses. DIoU incorporates the normalized distance between the centers of the predicted and the ground truth bounding boxes, showing much faster convergence in training concerning IoU and GIoU losses. CIoU considers in addition the aspect ratios of the predicted and target boxes, leading to still faster convergence and better performance. CIoU is defined by
\begin{equation}
    \mathcal{L}_{CIoU} = 1 - IoU + \frac{d^2(\mathbf{b}, \mathbf{b}^{gt})}{c^2} + \alpha v,
\end{equation}
\begin{equation}
    v = \frac{4}{\pi^2} \left( \arctan \frac{w^{gt}}{h^{gt}} - \arctan \frac{w}{h} \right)^2,
\end{equation}
\begin{equation}
    \alpha = \frac{v}{(1 - IoU) + v},
\end{equation}
% \frac{dist_2^2}{dist_C^2} + \frac{v^2}{(1 - IoU) + \nu}
where $d(\mathbf{b}, \mathbf{b}^{gt})$ is the Euclidean distance between the central points of bounding boxes $B$ and $B^{gt}$, $c$ is the diagonal length of the smallest enclosing box covering the two boxes, $v$ compares the predicted and target bounding boxes aspect ratios, and $\alpha > 0$ is a parameter giving higher relevance to the overlap area factor.

\section{Proposed Method}\label{sec:proposed_method}
\begin{figure*}[ht!]
    \centering
    \includegraphics[width=0.9\textwidth]{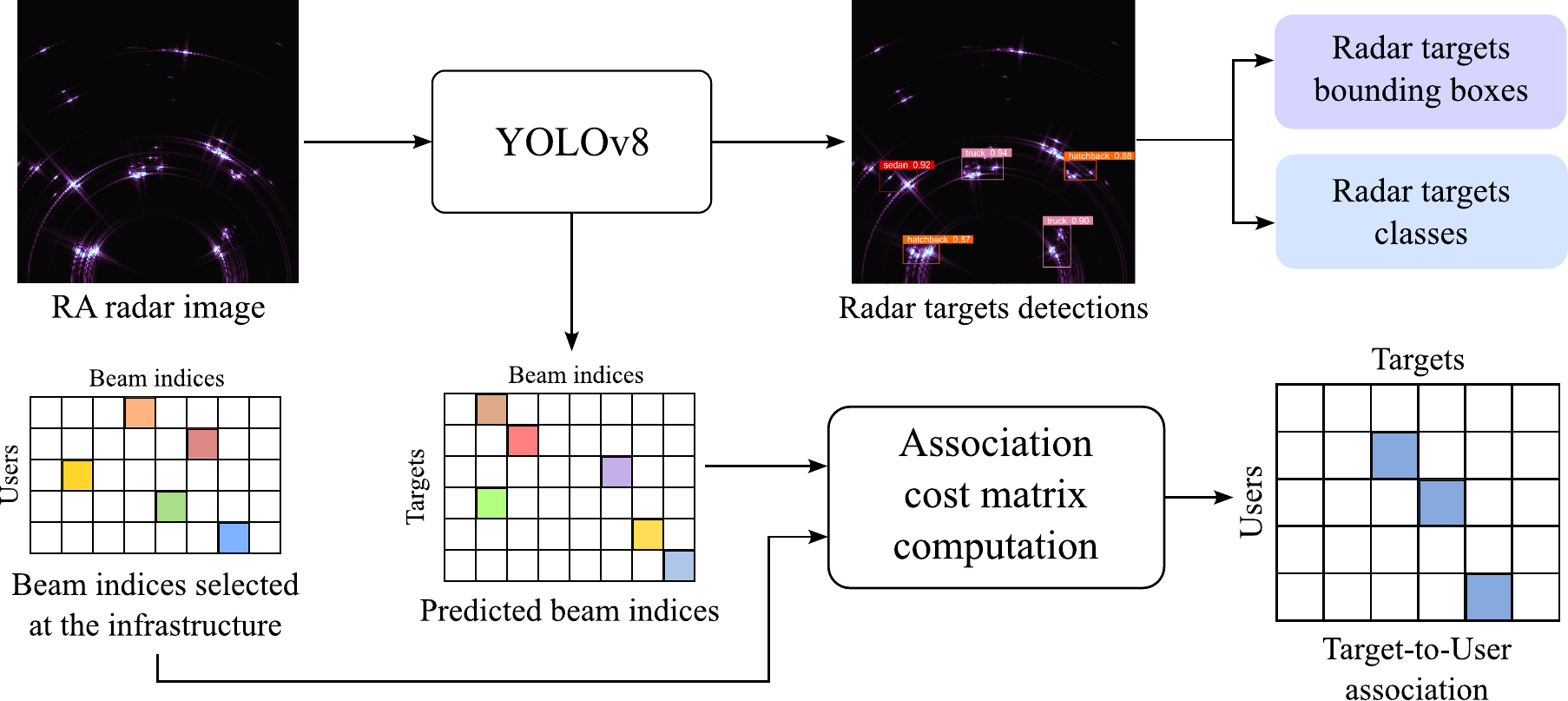}
    \caption{Proposed method for the integration of deep learning-based radar multi-target detection and beam prediction aimed at T2U association. A sequence of RA radar images is provided to a modified YOLOv8 model to jointly infer radar targets' positioning and classification information and the per-target beamforming vectors' classes. The predicted beam indices are then matched with the ones used at the BS to perform T2U association.}
    \label{fig:proposed_method}
\end{figure*}

In this section, we provide an overview of the proposed method for the integration of deep learning-based radar multi-target detection and beam prediction aimed at T2U association.
Motivated by the state-of-the-art performance of deep learning architectures for pattern recognition on complex structured data, we frame the radar multi-target detection and beam prediction tasks under a data-driven perspective, identifying in the YOLO model a consistent architectural bias to  flexibly integrate both tasks.
The proposed method is sketched in Fig. \ref{fig:proposed_method}. To clarify the aim of the depicted pipeline, we here briefly describe the proposed processing steps. The system inputs are radar range-angle (RA) image and a set of analog precoding beamforming vectors $\mathbf{f}$ selected at the BS for each vehicular target in the scene. The two inputs are assumed to be synchronized. In the following, we assume that the targets are vehicles crossing the scenario and that each vehicle is equipped with a VE as described in Section \ref{sec:communication_model}. Hence, we investigate the identification of radar targets within RA images, the distinguishability of the corresponding VEs in the beamspace, and the association of radar targets to VEs.
A description of the considered hybrid sub-connected communication architecture is provided in Section \ref{sec:communication_model}, along with a specification of the considered analog beamforming codebook. We refer the reader to Section \ref{sec:results} for a thorough description of the simulation framework, post-processing steps, and radar RA images generation.

For radar target detection, we consider the latest version of the YOLO deep learning model, i.e., YOLOv8 \cite{yolov8}, detailed in Section \ref{sec:YOLO_model}.
We integrated the YOLOv8 model to simultaneously infer targets' bounding boxes, classes, and analog beam indices, as detailed in Section \ref{sec:radar_target_detection}. To perform T2U association, the analog codebook beam indices inferred through YOLOv8 are compared with the beam indices observed during communication, as detailed in Section \ref{sec:t2u_association}.

\subsection{Deep Learning-based Joint Beam Prediction and Radar Multi-Target Detection}\label{sec:radar_target_detection}

\subsubsection{Radar multi-target detection}
We consider the problem of multi-target detection of vehicular radar targets over radar images. Range-Angle (RA) images have been shown in \cite{demirhan2022radar} to enhance beam prediction accuracy with respect to radar cube (i.e., tensors aggregating radar RA images at different velocity values) or Range-Velocity maps data, highly reducing computational complexity with respect to radar cube inputs owing to the considerably lower data dimensions. Therefore, in the following, we consider sequences of RA radar images as inputs to the YOLOv8 model.

We denote the sequence of contiguous input radar RA images by $I_{t}$, where $t = 1, \cdots, T$ indicates the sampling time step and $T$ is the total number of RA radar images in the sequence. Each vehicular target $k$ is represented by its bounding box $B_{t, k}$, and its class $c_{t, k} \in 1, \cdots, C$, with $C$ denoting the total number of target classes. We will refer to generic bounding boxes by omitting the time and object subscripts. In this paper, we consider three different vehicular target classes ($C = 3$), i.e., sedan, hatchback, and truck. A bounding box is described by $B_{t, k} = (x, y, w, h)$, where $(x, y)$ are the bounding box center coordinates, $w$ and $h$ are, respectively, the bounding box width and height within the image, and $k$ indexes an object in the radar image. Bounding box center position, width, and height are assumed to be normalized with respect to the radar image $I_{t}$ dimensions, i.e., $x, y, w, h \in [0, 1]$.

\subsubsection{Beam prediction}
We discretize the analog beam prediction problem as the inference of the beam index within a fixed-size beamforming codebook, whose features are specified in Section \ref{sec:communication_model}. Therefore, the analog beam prediction problem can be naturally defined as a multi-class classification problem over the set of indices of the vectors within the precoding analog beamforming codebook. We denote with $(f_{t, k}^{h}, f_{t, k}^{v})$ the indices within the horizontal and vertical beamforming codebooks of the beamforming vectors $(\mathbf{f}_{t, k}^{h}, \mathbf{f}_{t, k}^{v})$, used by the BS for analog precoding towards its VE, where $f_{t, k}^{h} \in 1, \cdots, N_T^{h}$, $f_{t, k}^{v} \in 1, \cdots, N_T^{v}$, and $N_T^{h}$, $N_T^{v}$ is as described in Section \ref{sec:communication_model}. In the following, we will refer to these indices as \textit{beam indices}.

In the YOLOv8 model, the classification head output is extended to contain the beam indices' classes besides the vehicular target classes. The beamforming codebook can be split into two separate codebooks over the horizontal and vertical antenna array directions. Therefore, we divide the beam prediction choice over the horizontal beam indices $\{f_{i}^{h}\}_{i=1, \cdots, N_T^{h}}$ and vertical beam indices $\{f_{j}^{v}\}_{j=1, \cdots, N_T^{v}}$. The resulting total number of additional beam prediction classes per object is 
\begin{equation}
    C_{beam} = N_T^{h} + N_T^{v}
\end{equation}
instead of $N_T^{h} \cdot N_T^{v}$ of the whole beamforming vectors when the array horizontal and vertical directions are jointly considered. This considerably reduces the computational effort by indexing each beamforming vector through the inferred horizontal and vertical coordinates, i.e.,
\begin{equation}
    \mathbf{f}_{\hat{i}, \hat{j}} = \mathbf{f}_{\hat{i}}^{h} \otimes \mathbf{f}_{\hat{j}}^{v},
\end{equation}
where $\hat{i} = \hat{f}^{h}$ and $\hat{j} = \hat{f}^{v}$ are, respectively, the inferred horizontal and vertical beam indices, and the analog precoding beamforming vector is provided by the Kronecker product of the beamforming vectors over the horizontal and vertical array directions, respectively.

We assume that a set of ground truth azimuth and elevation beam prediction labels are collected for training at the BS and that the matching between beam prediction labels and sensed radar targets is known. In the simulation results proposed in Section \ref{sec:results}, this is achieved by matching the ray-tracing wireless channel and radar sensing simulations. In a real deployment, the same result may be achieved by collecting synchronized wireless channels and positioning information from cooperative vehicles crossing the urban scenario in the neighborhood of the BS.

We consider the BCE loss in eq. \eqref{eq:BCE_loss} to perform optimization over both target classes and beam prediction classes. Then, the total number of classes becomes $C = C_{target} + C_{beam}$. The most relevant target classes and beam prediction classes are separately determined by applying a threshold $\gamma_{class}$ on the inferred classification scores.

We notice that, with respect to position-based methods proposed in the literature, e.g., \cite{khan2020position}, the proposed method based on the YOLOv8 architecture end-to-end optimizes the beam prediction task considering the whole radar image as conditioning input and, therefore, the joint knowledge of the objects present in the image. Then, both global information on relative positioning of the perceived targets and local details on their state are available for the inference of the beam indices at the BS. This provides the possibility to take into account also anomalous conditions---e.g., dynamic blockage---that may depend on relative vehicular motion patterns, or to predict the availability of relay communication links that can be used when no line-of-sight (LoS) links are available.

\subsection{Beamspace T2U Association}\label{sec:t2u_association}
We propose here to associate radar targets to VEs in the beamspace leveraging the knowledge of the precoding beam indices selected for communication at the BS and the per-target beam indices inferred through the YOLOv8 architecture described in Section \ref{sec:radar_target_detection}.

Considering a fixed observation time step, the selected horizontal and vertical beam indices for a VE are denoted, respectively, by one-hot encoded vectors $\textbf{y}^{h} \in \{0, 1\}^{N_T^{h}}$, and $\textbf{y}^{v} \in \{0, 1\}^{N_T^{v}}$. The beam indices $\hat{\mathbf{y}}^{h} \in \mathbb{R}^{N_T^{h}}$ and $\hat{\mathbf{y}}^{v} \in \mathbb{R}^{N_T^{v}}$ represent the beam prediction logits output from the YOLOv8 classification head for the detected radar targets. The classification scores are processed by a softmax function
\begin{equation}
    \sigma(\mathbf{z})_i = \frac{e^{\mathbf{z}_i}}{\sum_{j=1}^{N} e^{\mathbf{z}_j}}
\end{equation}
to transform the output logits into discrete probability distributions over the horizontal and vertical codebook beam indices. The selected beam indices and the predicted ones are then compared by means of categorical cross-entropy:
\begin{equation}
    \mathcal{L}_{CCE} = -\sum_{i = 1}^{N_T^{h}} \mathbf{y}_i^{h} \log \sigma(\hat{\mathbf{y}}_i^{h}) - \sum_{j = 1}^{N_T^{v}} \mathbf{y}_j^{v} \log \sigma(\hat{\mathbf{y}}_j^{v}),
\end{equation}
where the softmax function $\sigma$ in the two sums is computed over $N_T^{h}$ and $N_T^{v}$ classes, respectively.

In this context, the BCE loss is used as a metric to build the association cost matrix $C^{T2U}$. The resulting assignment problem is represented by the following linear minimization problem:
\begin{equation}\label{eq:linear_assignment}
\min_{X} \sum_{k = 1}^K \sum_{v = 1}^V C_{k, v}^{T2U} X_{k, v},\\
\end{equation}
where $X_{k, v} \in \{0, 1\}$ represents an element of the assignment matrix $X \in \mathbb{R}^{K \times V}$, $K$ is the total number of detected radar targets, and $V$ is the total number of active VEs at the observation time step. The linear assignment problem \eqref{eq:linear_assignment} is then solved by means of the Hungarian algorithm.

The performance of T2U association is evaluated in terms of the probability of correct association between radar targets and VEs, provided by the ratio of correctly performed associations over the total number of targets in the scene averaged over simulation time steps.

\section{Simulation Results}\label{sec:results}
In this section, we present the simulation results achieved by the proposed methods. After introducing the utilized V2X channel simulation and radar imaging framework, we present the results achieved over the tasks of radar multi-target detection, beam prediction, and beamspace T2U association.

\begin{table}[t!]
\centering
\caption{Simulation parameters}
\begin{tabular}{r l c}
\toprule
\textbf{Task} & \textbf{Parameter} & \textbf{Value}\\
\noalign{\smallskip}
\hline
\noalign{\smallskip}

% Communication params
\multirow{5}{*}{\shortstack[r]{Ray-tracing\\ communication\\ channel}} & BS carrier frequency & $28$ GHz\\\noalign{\smallskip}
& Communication bandwidth & $100$ MHz\\\noalign{\smallskip}
& BS height from the ground & $5$ m\\\noalign{\smallskip}
& VEs height from vehicles' rooftops & $0.2$ m\\\noalign{\smallskip}
& Max. reflection, diffraction, scattering & $(7, 6, 1)$\\\noalign{\smallskip}
& Number of ray-tracing paths & $25$\\
\midrule

% Radar sensing params
\multirow{7}{*}{\shortstack[r]{Ray-tracing\\ sensing\\ channel}} & Radar carrier frequency & $28$ GHz\\\noalign{\smallskip}
& Radar bandwidth & $800$ MHz\\\noalign{\smallskip}
& Radar height from the ground & $5$ m\\\noalign{\smallskip}
& Max. reflection, diffraction & $(2, 1)$\\\noalign{\smallskip}
& Number of ray-tracing paths & $2\cdot 10^5$\\\noalign{\smallskip}
& Number of azimuth antennas & $128$\\\noalign{\smallskip}
& Number of elevation antennas & $4$\\

\bottomrule
\end{tabular}
\label{tab:simulation_parameters}
\end{table}

For simulations in vehicular environments, we selected the three urban scenarios depicted in Fig. \ref{fig:simulation_scenarios}. Below each scenario, we report a RA radar imaging sample acquired during vehicles motion. The chosen scenarios lead to different traffic vehicular patterns: in Scenario A, the ISAC system laterally observes a single road bidirectionally crossed by multiple vehicles; Scenario B contains a roundabout, which includes rotational motion patterns that are not present in the simplified Scenario A; Scenario C contains both intersections and curved trajectories, resulting in different vehicle motion patterns with respect to Scenarios A and B.

\subsection{Simulation framework}\label{sec:results_simulation}
\begin{figure}[t!]
    \centering
    \includegraphics[width=0.9\columnwidth]{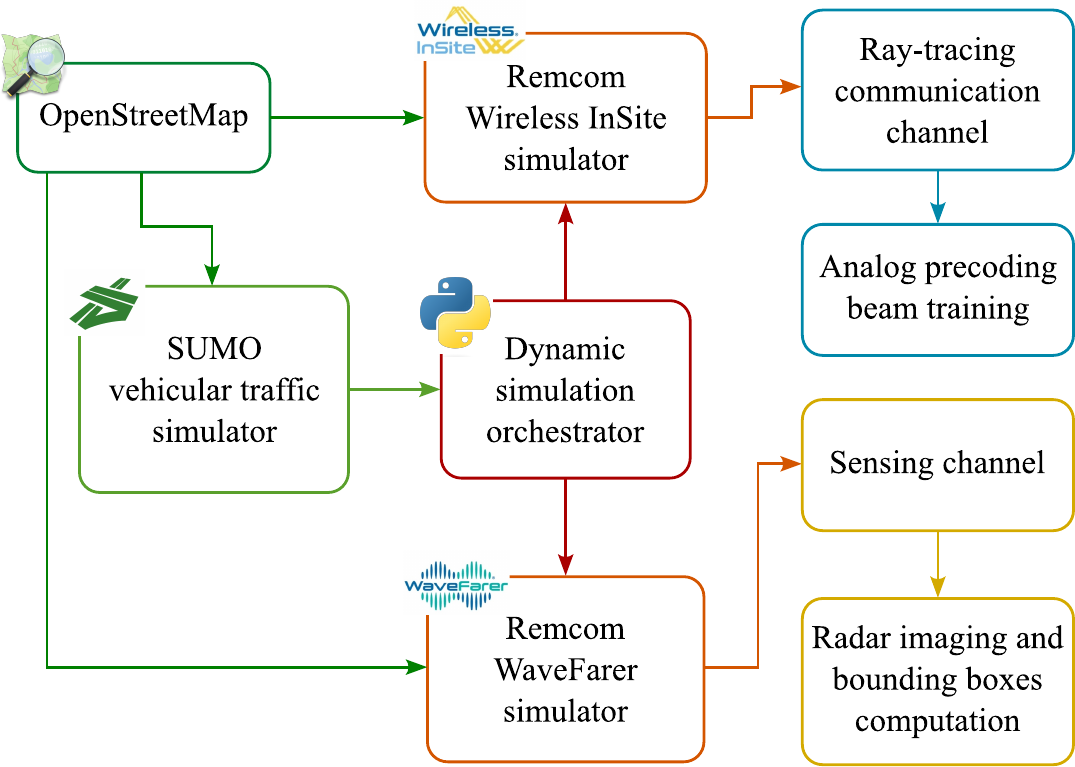}
    \caption{Communication and sensing channel simulation framework.}
    \label{fig:simulation_framework}
\end{figure}

The developed simulation framework is depicted in Fig. \ref{fig:simulation_framework}, which highlights the utilized simulation software and its use in relation to the tasks examined in this paper. Table \ref{tab:simulation_parameters} shows the simulation parameters selected for the three chosen scenarios. We detail here the simulation workflow depicted in Fig. \ref{fig:simulation_framework}.

\begin{figure*}[t!]
    \centering
    \includegraphics[width=.9\textwidth]{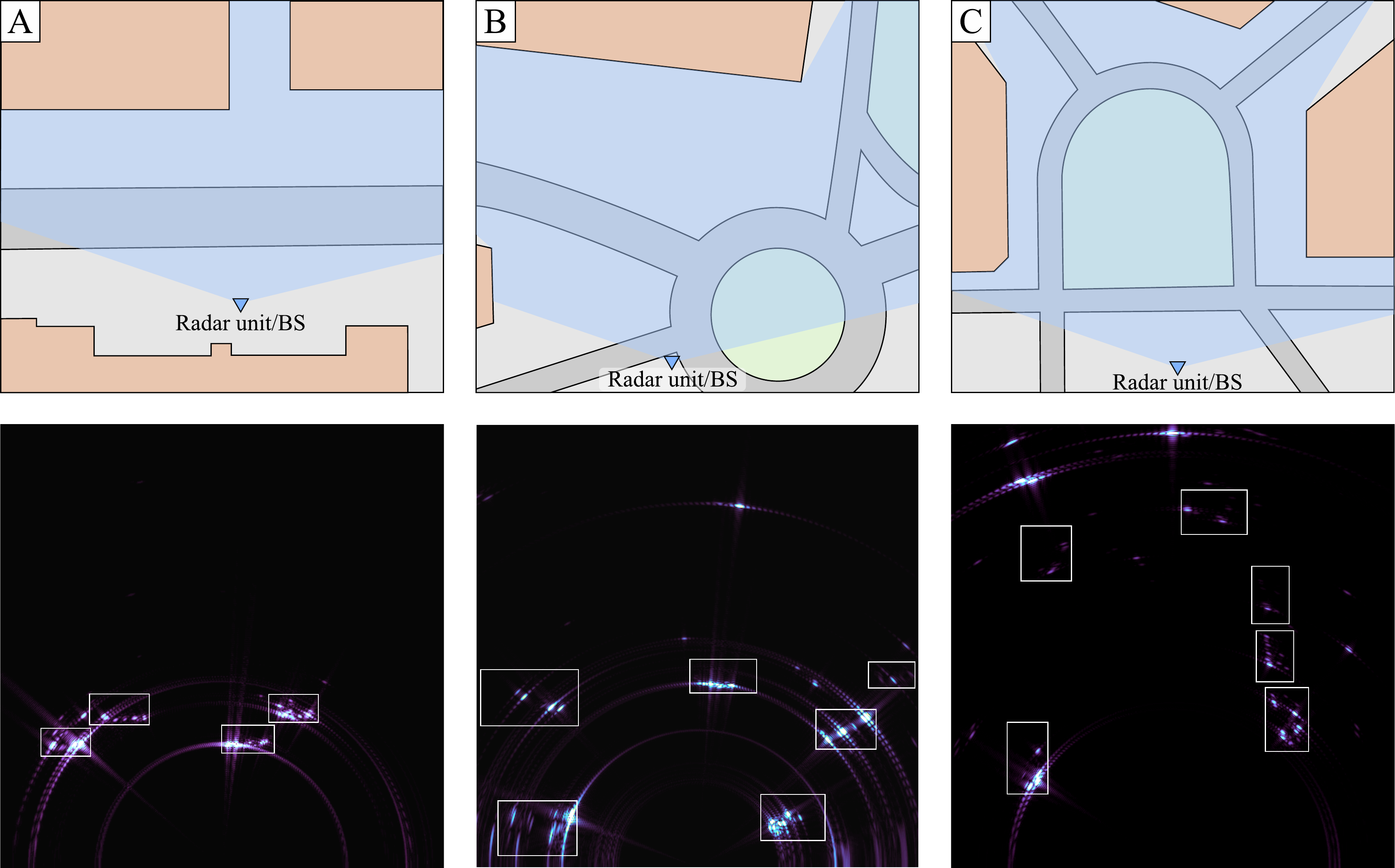}
    \caption{Considered urban scenarios (top row) with FOV of the radar sensor---represented as a shaded area,---and radar image samples (bottom row) showing vehicular targets crossing the scenario. Bounding boxes indicate the vehicles' positions and extensions within the radar images.}
    \label{fig:simulation_scenarios}
\end{figure*}

\textbf{Data acquisition and pre-processing}: Building structures and road traffic networks data are exported from OpenStreetMap\cite{openstreetmap} for the selected urban scenarios depicted in Fig. \ref{fig:simulation_scenarios}. The collected map data are post-processed through the Blender \cite{blender} 3D modeling software and the Blender-OSM \cite{blenderosm} plugin to extract the buildings' 3D structures required for wireless propagation simulations. We considered three different vehicle types, i.e., sedan, hatchback, and truck, which have varied geometric features and impact differently on wireless propagation.

\textbf{Vehicular traffic simulation}: Vehicular traffic networks are extracted from the OpenStreetMap data and a set of realistic vehicular trajectories is generated for each scenario by means of the Simulation of Urban Mobility (SUMO) microscopic vehicular traffic simulator \cite{sumo}. Random generation of vehicular traffic has been defined by selecting the vehicles' density per lane km to obtain a regular traffic flow through the networks. Vehicles have been generated at the fringe of the maps and crossed the scenarios according to randomly generated routes. For each discrete simulation time step $t$, and for each simulated vehicle at time $t$, we collected information on the state of the vehicle, i.e., position, heading angle, and speed. Vehicular traffic simulations have been performed using a simulation time step of $0.1$ s, with a simulation duration of $60$ s for each considered scenario.

\textbf{Communication channel simulation}: The mmWave communication channel has been simulated using Remcom Wireless InSite \cite{wireless_insite}. Owing to channel sparsity in the space and time domains, ray-tracing is one of the most accurate techniques for the simulation of propagation environments at high frequencies \cite{degli2021ray}. Indeed, electromagnetic (EM) propagation at mmWaves strictly depends on the geometrical features of the environment and is widely explained by a few different types of physical interactions with the environment, including reflection, refraction, diffraction, transmission, and scattering.

Frame-by-frame dynamic simulation is achieved by changing the position and orientation of the vehicles' 3D meshes within the Wireless InSite 3D environment according to the vehicular trajectories generated by SUMO. We consider a VE on the rooftop of each simulated vehicle while the BS is positioned as represented in Fig. \ref{fig:simulation_scenarios}. Isotropic antennas are set for both the BS and the VEs to collect the rays from the BS to the VEs over the simulation time steps. For each VE and for each simulated ray-tracing path, we collect from Remcom Wireless InSite the direction of departure (DoD), direction of arrival (DoA), received power, phase, and delay. We report the communication channel simulation parameters in Table \ref{tab:simulation_parameters}.

\textbf{Analog precoding beam training}: The analog beamforming vectors to be selected for communication, among the ones in the precoding codebook, are determined through a beam training procedure exhaustively spanning the combinations of Tx and Rx beams. During simulations, VEs are equipped with a $2\times2$ rectangular antenna array lying on the vehicle rooftop, while rectangular antenna arrays from $2\times2$ to $64\times64$ antennas are considered at the BS. Beam training is performed in different signal-to-noise ratio (SNR) per antenna conditions, starting from -55 dB to -10 dB. The YOLO-based beam prediction method proposed in Section \ref{sec:proposed_method} is trained over analog precoding beam indices determined at -10 dB SNR per antenna, while the performance of the model is examined over all the considered communication SNRs range.

% \textcolor{red}{unire a sotto e snellire} 
\textbf{Radar sensing channel simulation}: The mmWave radar sensing channel has been simulated by means of Remcom WaveFarer \cite{wavefarer}. Differently from Remcom Wireless InSite---which simulates the leading propagation paths for communication channel modeling purposes,---Remcom WaveFarer generates a high number of propagation paths to achieve the resolution required for accurate radar sensor simulation at high frequencies. As for the communication channel simulation, we perform a frame-by-frame ray-tracing simulation moving vehicles according to the trajectories generated by means of SUMO. Separate EM materials are assigned to different vehicle parts to increase the radar sensing simulation accuracy (i.e., PEC for vehicle body and rims, glass for windows, and rubber for tires). Besides direction of departure (DoD), direction of arrival (DoA), received power, phase, and delay, we collect from Remcom WaveFarer the positions of the interaction points of the propagation rays with the environment. The sensing channel simulation parameters are reported in Table \ref{tab:simulation_parameters}.

\textbf{Radar targets' bounding boxes computation}: Since the radar sensing channel can be enriched by a variety of interactions with the environment (e.g., owing to reflections and diffractions), it is crucial to effectively distinguish a radar target from the environment and provide a suitable definition of the object's boundaries. In a radar image, this task is not as immediate as for camera images owing to the different image formation process. Indeed, the transmitted radar signal can undergo multiple interactions with the propagation environment, resulting in distortions of target images. We employ a heuristic approach to determine these boundaries exploiting the availability of detailed propagation information from the Remcom WaveFarer sensing channel simulator. Therefore, we divide the bounding boxes computation into three steps:
\begin{enumerate}
    \item Identification of the interaction points lying on the target or strictly related to the latter, achieved evaluating the orthogonal distance of the wireless propagation interaction points with respect to the vehicle oriented 3D mesh and selecting the points belonging to the related propagation rays (derived from the WaveFarer ray-tracing procedure).
    \item Filtering of the identified points by power and by target distance, where the filtering is applied to take into consideration interaction points near to the target that are sufficiently important (e.g., near-to-target ground reflections); this phase allows to include points that can be relevant for the detection training process but do not extend the bounding box to non-representative target dimensions.
    \item Spherical projection onto the radar slant-range plane of the filtered points and generation of the bounding box enclosing the projected points.
\end{enumerate}

\subsection{Radar Multi-Target Detection and Beam Prediction}

We present here the results obtained performing radar multi-target detection by means of YOLOv8 using 5-fold cross-validation. Results are averaged among the three considered scenarios depicted in Fig. \ref{fig:simulation_scenarios}. Radar target detection has been performed from $512 \times 512$ RA radar images. As described in Section \ref{sec:radar_target_detection}, we jointly perform radar target detection and analog beam prediction for each detected target. We selected YOLOv8 \textit{small} model size (YOLOv8s) as a trade-off between detection/classification performance and real-time efficiency. The fixed-size model performance is examined over a selection of rising planar array dimensions at the BS, resulting in increasing beamforming resolution and number of beam indices for classification. We trained the YOLOv8s models on a Linux server accelerated by NVIDIA RTX 6000 GPUs. Training has been performed according to the hyperparameters specified in Table \ref{tab:hyperparameters}. Early stopping after a patience window of $50$ epochs has been considered. The trained models converged activating the early stopping condition within $300$ epochs.

Table \ref{tab:target_detection_performance} reports the detection and classification performance of the trained YOLOv8 model for different rectangular antenna array sizes at the BS. The array sizes are indicated in the form $N_T^{h} \times N_T^{v}$. The antenna array size determines the size of the codebooks for analog precoding along the horizontal and vertical antenna array dimensions, as reported in Section \ref{sec:communication_model}. The classification performance is shown in terms of the precision, recall, and F1 metrics. Detection performance is measured by mean average precision with an IoU threshold of $0.5$ (mAP\textsubscript{50}), and by the mean of mean average precision mAP\textsubscript{50-95} determined by averaging mAP at $10$ different IoU thresholds between $0.5$ and $0.95$ (extremes included). The detection and classification performance is provided over both the radar targets classification task (i.e., whether the vehicular targets belong to the \textit{sedan}, \textit{hatchback}, or \textit{truck} classes) and on beam prediction.

\begin{table*}[ht!]
\centering
\caption{Mutli-target detection performance on the three considered scenarios (5-fold cross-validation) for a selection of antenna array sizes at the BS. Precision, recall and F1-score provide the classification performance, while mAP\textsubscript{50} and mAP\textsubscript{50-95} account for both classification and detection. The metrics are separately evaluated on the targets classification and beam prediction tasks.}
\begin{tabular}{c | c c c c c | c c c c c}
\multicolumn{1}{c}{} & \multicolumn{5}{c}{Targets classification} & \multicolumn{5}{c}{Beam prediction}\\\noalign{\smallskip}
\toprule
\textbf{Array size} & \textbf{Precision} & \textbf{Recall} & \textbf{F1} & \textbf{mAP\textsubscript{50}} & \textbf{mAP\textsubscript{50-95}} & \textbf{Precision} & \textbf{Recall} & \textbf{F1} & \textbf{mAP\textsubscript{50}} & \textbf{mAP\textsubscript{50-95}}\\
\noalign{\smallskip}
\hline
\noalign{\smallskip}
$2\times2$ & 0.90 & 0.78 & 0.80 & 0.92 & 0.79 & 0.91 & 0.79 & 0.81 & 0.92 & 0.79\\
$4\times4$ & 0.90 & 0.76 & 0.78 & 0.90 & 0.76 & 0.84 & 0.70 & 0.69 & 0.86 & 0.73\\
$8\times8$ & 0.89 & 0.77 & 0.79 & 0.89 & 0.76 & 0.80 & 0.70 & 0.69 & 0.84 & 0.71\\
$16\times16$ & 0.88 & 0.77 & 0.79 & 0.89 & 0.76 & 0.76 & 0.65 & 0.63 & 0.79 & 0.67\\
$32\times32$ & 0.89 & 0.77 & 0.79 & 0.90 & 0.77 & 0.73 & 0.61 & 0.58 & 0.74 & 0.64\\
$64\times64$ & 0.88 & 0.79 & 0.81 & 0.89 & 0.77 & 0.69 & 0.55 & 0.51 & 0.67 & 0.59\\
\bottomrule
\end{tabular}
\label{tab:target_detection_performance}
\end{table*}

\begin{figure*}[ht!]
    \centering
    \subfloat[][]{\includegraphics[width=.25\textwidth]{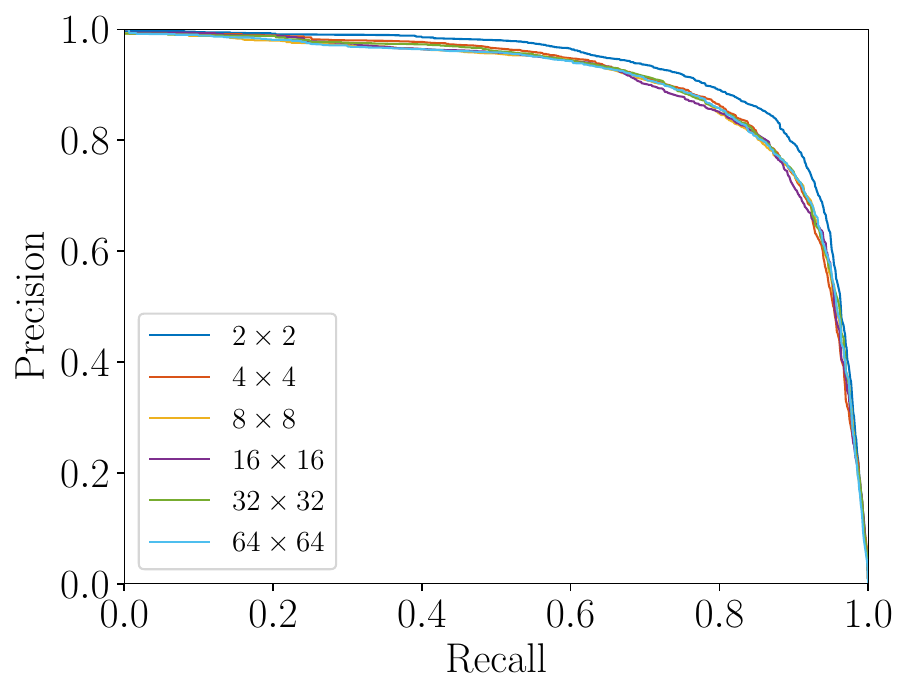}}
    \subfloat[][]{\includegraphics[width=.25\textwidth]{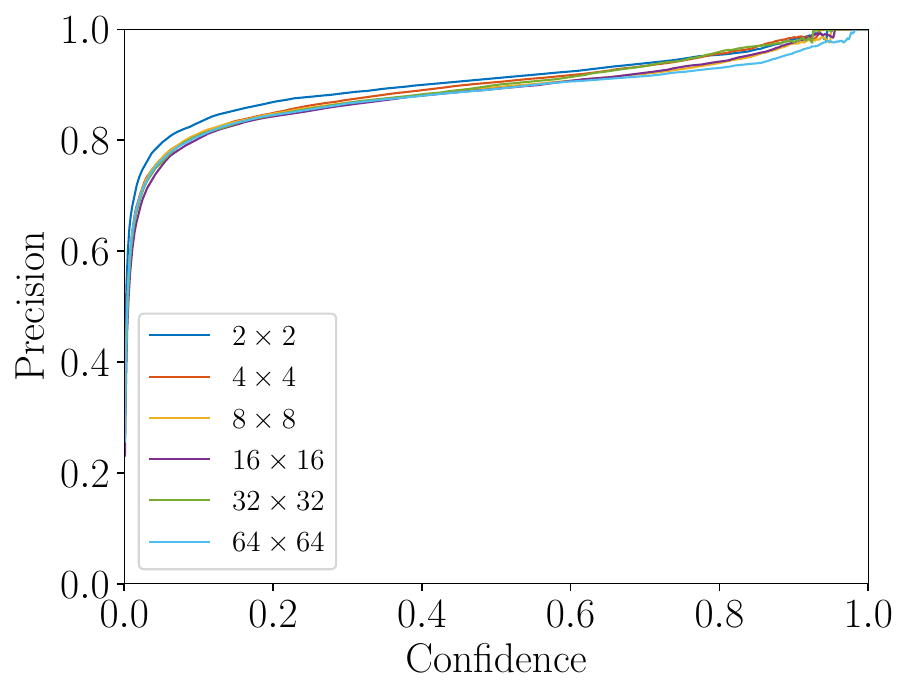}}
    \subfloat[][]{\includegraphics[width=.25\textwidth]{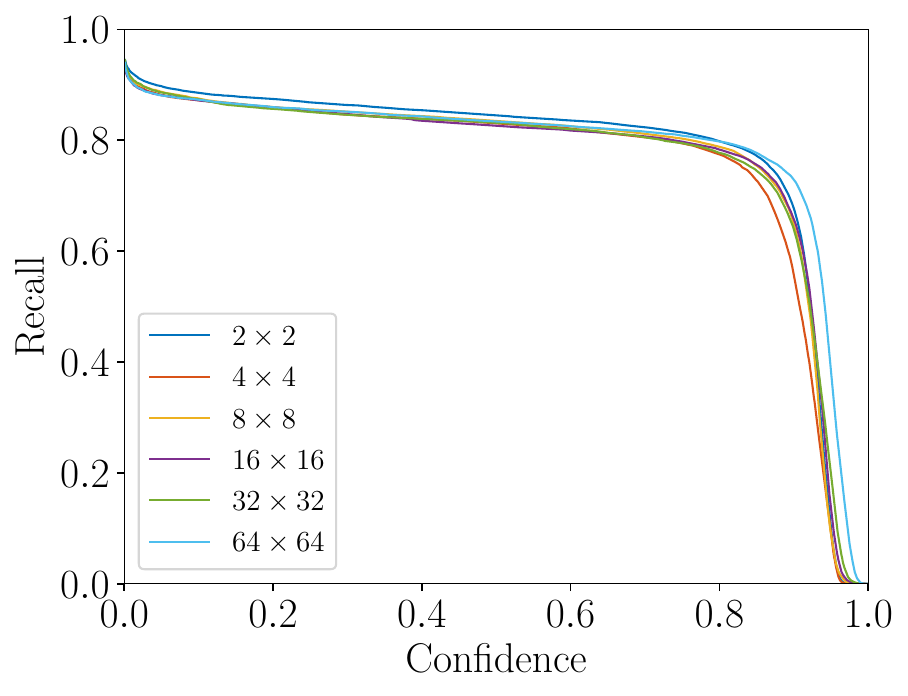}}
    \subfloat[][]{\includegraphics[width=.25\textwidth]{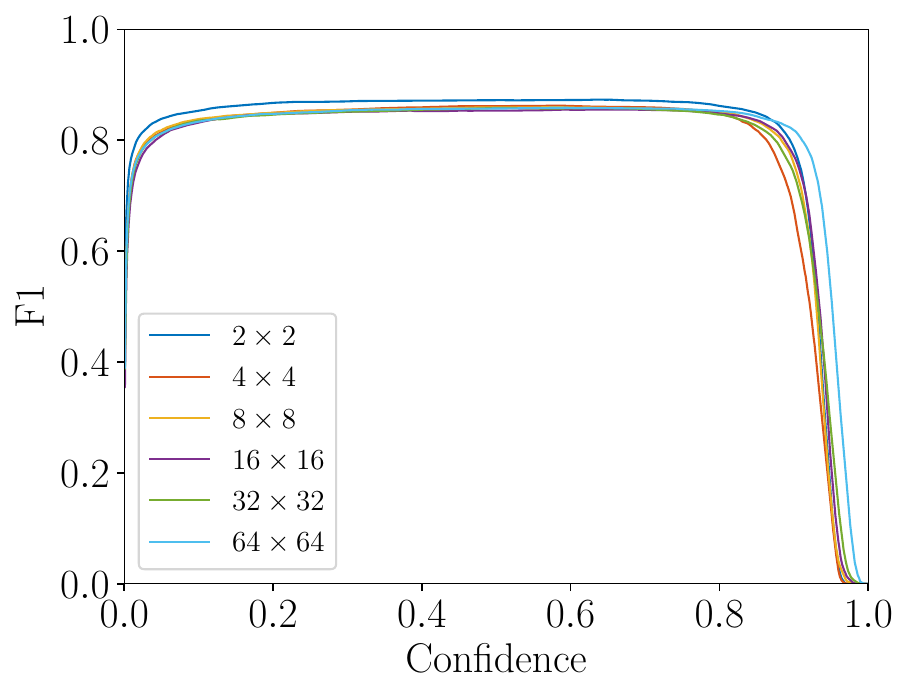}}\\
    \subfloat[][]{\includegraphics[width=.25\textwidth]{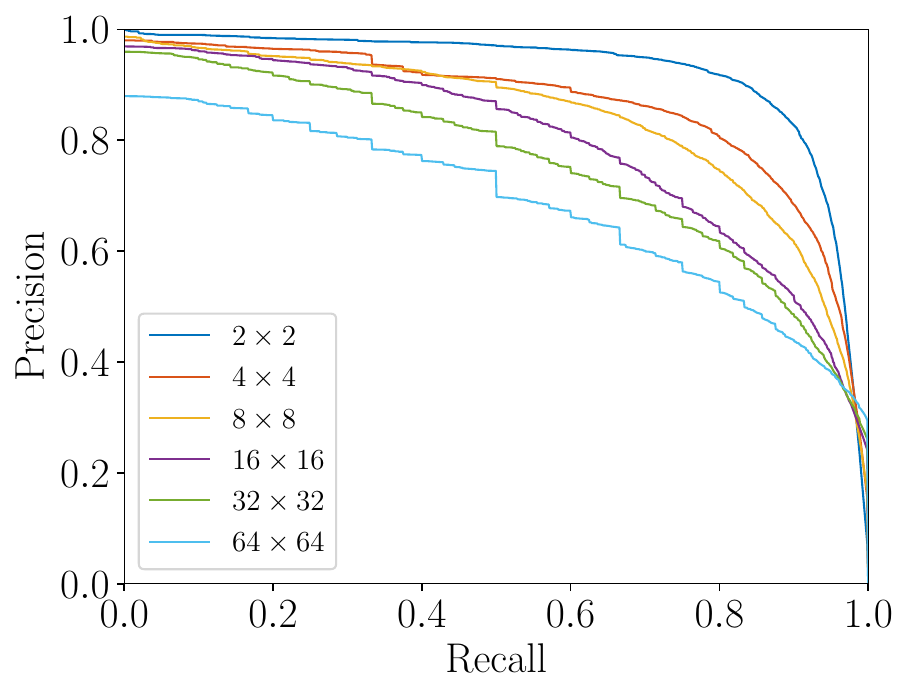}}
    \subfloat[][]{\includegraphics[width=.25\textwidth]{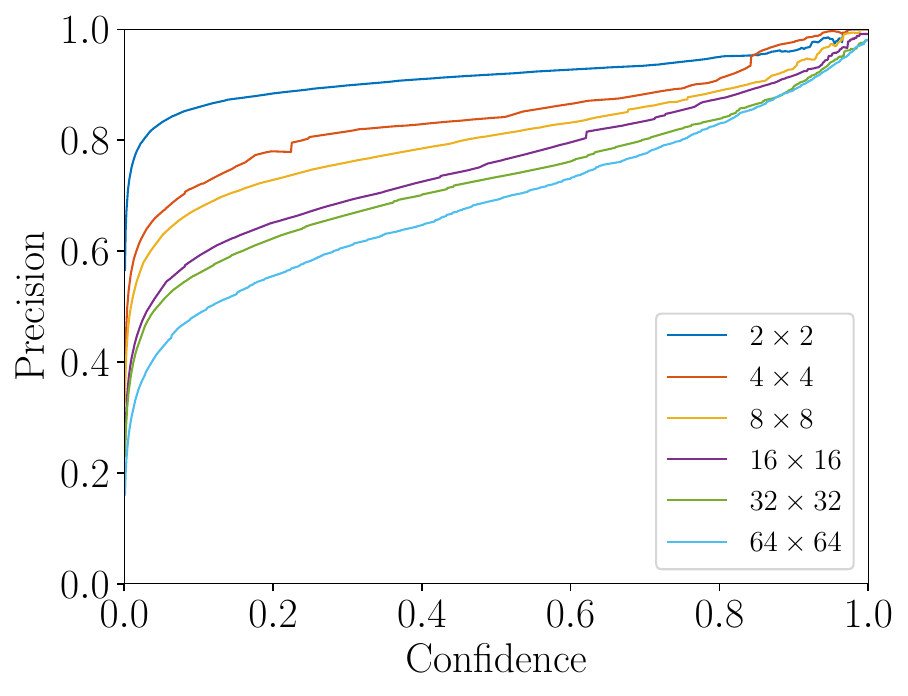}}
    \subfloat[][]{\includegraphics[width=.25\textwidth]{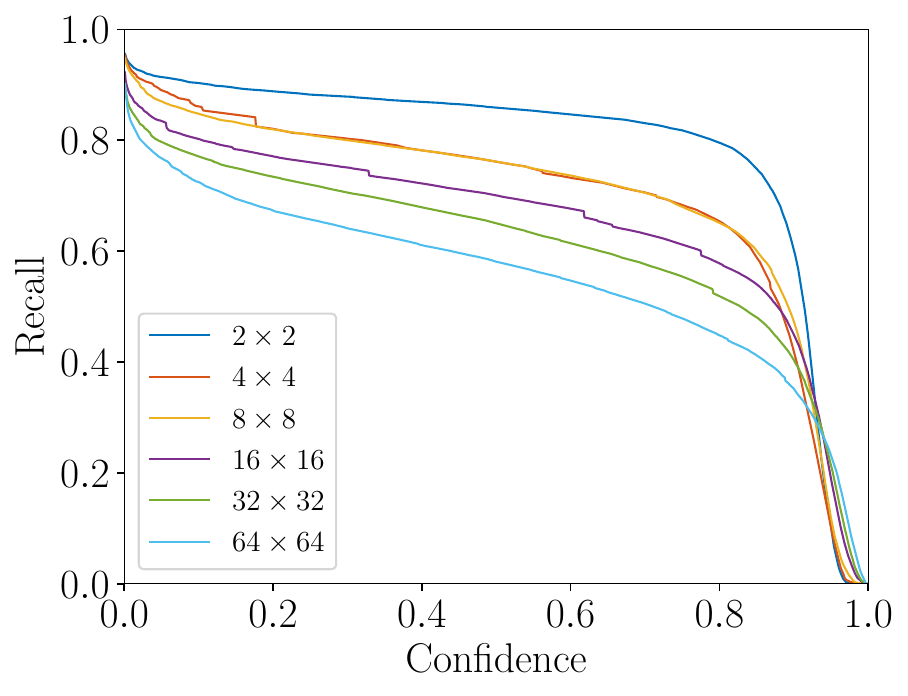}}
    \subfloat[][]{\includegraphics[width=.25\textwidth]{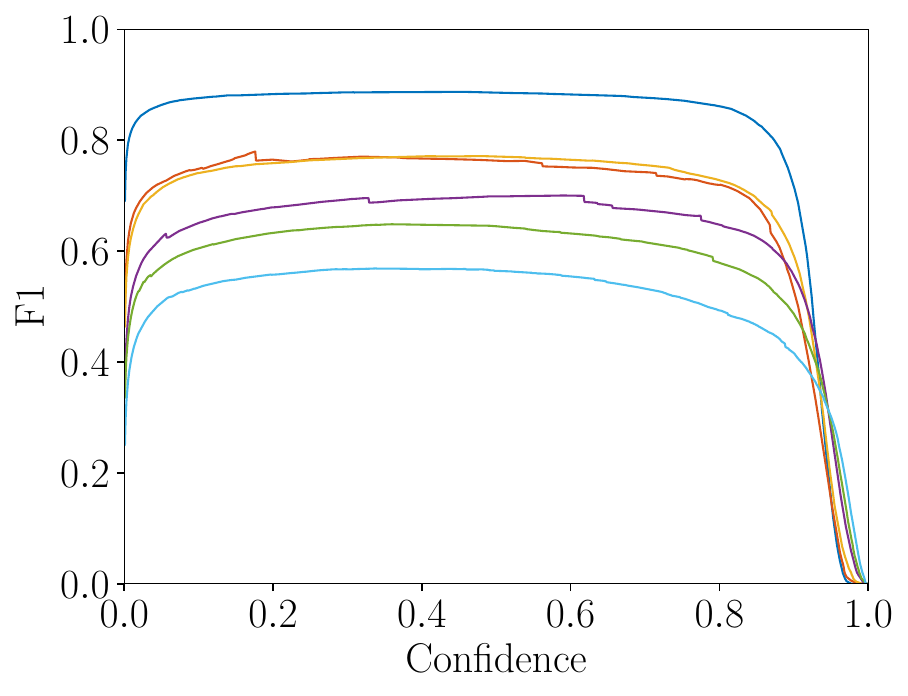}}\\
    \caption{YOLOv8 detection and classification 5-fold cross-validation performance over radar target classification (top row) and beam prediction (bottom row). Results are provided in terms of precision/recall, precision/confidence, recall/confidence and F1-score/confidence curves for a selection of antenna array sizes at the BS.}
    \label{fig:target_detection_results}
\end{figure*}

\begin{table}[t!]
\centering
\caption{Comparison of the YOLOv8s, Faster R-CNN and SSD object detection models over scenario C (Fig. \ref{fig:simulation_scenarios}).}
\begin{tabular}{r c c c c}
\toprule
\textbf{Detection model} & \textbf{Precision} & \textbf{Recall} & \textbf{mAP\textsubscript{50}} & \textbf{\textbf{mAP\textsubscript{50-95}}}\\
\noalign{\smallskip}
\hline
\noalign{\smallskip}
% YOLO comparison
YOLOv8s & 0.96 &  0.77 & 0.89 & 0.70\\\noalign{\smallskip}
Faster R-CNN & 0.83 & 0.88 & 0.82 & 0.66\\\noalign{\smallskip}
SSD & 0.79 & 0.90 & 0.79 & 0.53\\\noalign{\smallskip}
\bottomrule
\end{tabular}

\label{tab:yolo_comparison}
\end{table}

We provide in Table \ref{tab:yolo_comparison} a performance comparison of the selected YOLOv8s model with two widespread deep learning-based object detection models, i.e., Faster R-CNN \cite{ren2015faster} and SSD \cite{liu2016ssd}, over the radar vehicular target detection and classification task. We select a ResNet101\cite{he2016deep} pre-trained backbone and an FPN neck for the Faster R-CNN model, while the SSD model is equipped with a pre-trained VGG16 \cite{simonyan2014very} backbone. We notice that, while Faster R-CNN is a two-stage detector---which separates the regions proposal phase from their classification,---both YOLOv8 and SSD are single stage detectors. The three methods are compared on the scenario C depicted in Fig. \ref{fig:simulation_scenarios}, which is the one exhibiting the most varied vehicular patterns among the simulated environments. The hyperparameters have been manually tuned to achieve high detection performance. Setting the maximum number of training epochs to 300, the highest mAP validation performance has been achieved at 56 epochs for YOLOv8s, 30 epochs for Faster R-CNN and 99 epochs for SSD. The training and validation of the two methods over the radar images dataset has been performed by means of the MMDetection benchmarking library\cite{mmdetection}. We refer the reader to \cite{ren2015faster} and \cite{liu2016ssd} for a thorough description of the Faster R-CNN and SSD architectures.

While providing notable detection and classification performance over the considered scenario, YOLOv8s presents a higher computational efficiency and a considerably lower number of model parameters with respect to both the considered Faster R-CNN and SSD architectures. The selected configurations lead to $\approx11.2 \cdot 10^6$ model parameters for YOLOv8s, $\approx60.7 \cdot 10^6$ for Faster R-CNN, and $\approx24.7 \cdot 10^6$ for SSD. Moreover, Faster R-CNN showed a higher detection performance with respect to SSD. This behaviour of Faster R-CNN and SSD over radar images is similar to the one obtained in \cite{stroescu2021object}, where the authors show a comparable gap between SSD and Faster R-CNN for training over real-world acquired data. Nevertheless, we notice that this behavior can depend on the choice of several model features, e.g., the backbone type or the radar image pre-processing steps, and that, to the best of our knowledge, a thorough architectural comparison on the radar target detection task over the available deep learning object detection models is missing in the specific literature. The proposed comparison aims to show that the radar target detection task can be effectively learned by diverse deep learning-based object detection architectures originally designed (and partially pre-trained) for camera data.

We notice that the performance under radar target classification is minimally affected by the number of additional classes (i.e., $N_T^{h} + N_T^{v}$) owing to the beam prediction task, with a $\approx 2.5 \%$ loss in mAP\textsubscript{50-95} between the $2 \times 2$ and the $64 \times 64$ antenna array size cases. We show that the classification and detection metrics over horizontal and vertical beams prediction depend instead on the beamforming codebook size, resulting in the highest performance for the $2 \times 2$ antenna array case up to $\approx 25 \%$ loss in mAP\textsubscript{50-95} for a $64 \times 64$ antenna array.

\begin{figure}
    \centering
    \includegraphics[width=0.95\columnwidth]{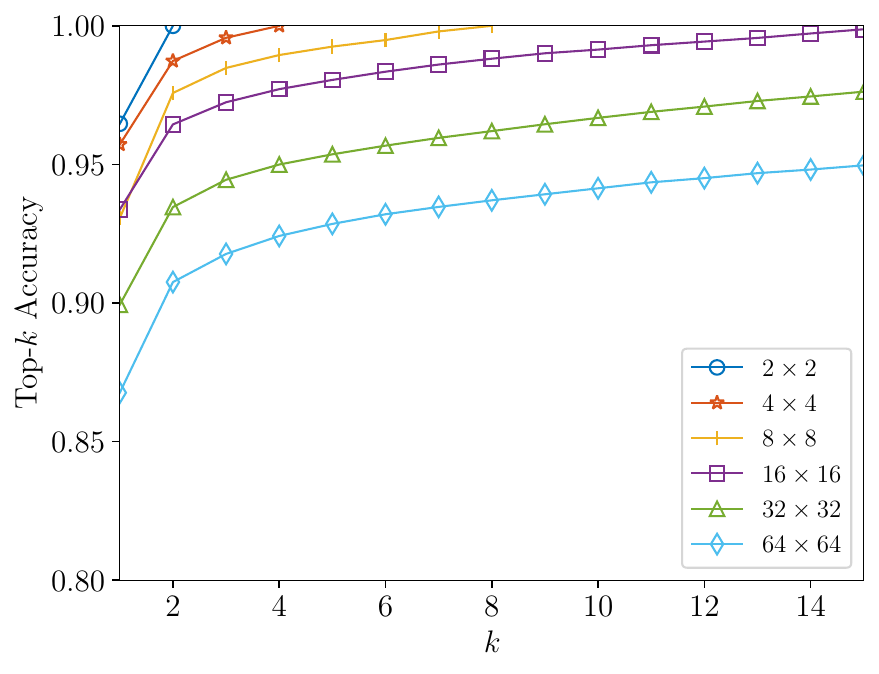}
    \caption{Beam prediction top-$k$ accuracy over different rectangular antenna array dimensions at the BS.}
    \label{fig:beam_prediction_accuracy}
\end{figure}

Fig. \ref{fig:target_detection_results} presents the YOLOv8s performance by means of precision-recall (PR), precision-confidence (PC), recall-confidence (RC), and F1-confidence (FC) graphs for both the target classification (Figs. \ref{fig:target_detection_results}a to \ref{fig:target_detection_results}d) and beam prediction (Figs. \ref{fig:target_detection_results}e to \ref{fig:target_detection_results}h) tasks. The YOLO model performance is summarily described by the area under the proposed curves. The results reflect the same behavior presented in Table \ref{tab:target_detection_performance}. Indeed the metrics related to the targets classification task are almost unaffected by the precoding codebook size, whereas the metrics evaluated on the beam prediction task show decreasing performance owing to the increase in the output inference space over the same YOLOv8 model size.

The top-$k$ classification accuracy for the beam prediction task is provided in Fig. \ref{fig:beam_prediction_accuracy}. The classification accuracy is evaluated as the average between the horizontal and the vertical beam classification top-$k$ accuracy. The proposed model shows significant top-$1$ accuracy over the considered antenna array dimensions at the BS, with a $\approx 9.4\%$ performance loss for a $64 \times 64$ antenna array at the BS with respect to a $2 \times 2$ antenna array. Moreover, the model presents a similar top-$k$ accuracy increase over the all the considered antenna array sizes, showing an average enhancement of $\approx 5.7 \%$ from top-$1$ to top-$5$ accuracy over $64 \times 64$ antenna arrays.

\begin{table}[t!]
\centering
\caption{Model hyperparameters.}
\begin{tabular}{r l c}
\toprule
\textbf{Model} & \textbf{Hyperparameter} & \textbf{Value}\\
\noalign{\smallskip}
\hline
\noalign{\smallskip}

% Target detection - YOLOv8
\multirow{7}{*}{\shortstack[r]{Radar\\ multi-target\\ detection\\ (YOLOv8)}} & Input images size & $512 \times 512$\\\noalign{\smallskip}
& Max. number of epochs & $500$\\\noalign{\smallskip}
& Max. number of detections & $300$\\\noalign{\smallskip}
& Confidence threshold & $0.25$\\\noalign{\smallskip}
& IoU threshold & $0.7$\\\noalign{\smallskip}
& Batch size & $16$\\\noalign{\smallskip}
& Optimizer & SGD\\

\bottomrule
\end{tabular}
\label{tab:hyperparameters}
\end{table}

\subsection{T2U Association}
We present here the results achieved applying the T2U association procedure described in Section \ref{sec:t2u_association} to match the detected radar targets to VEs in the beamspace. Radar target detection and per-target beam prediction are performed by means of the YOLO architecture proposed in Section \ref{sec:radar_target_detection} according to the hyperparameters specified in Table \ref{tab:hyperparameters}. Analog precoding beam training is performed at the BS by an exhaustive span of the Tx. and Rx. beamforming codebooks at each simulation time step, as described in Section \ref{sec:results_simulation}. We assume that the vehicular targets can be sensed by the BS and that each target is equipped with a VE.

\begin{figure}
    \centering
    \includegraphics[width=0.95\columnwidth]{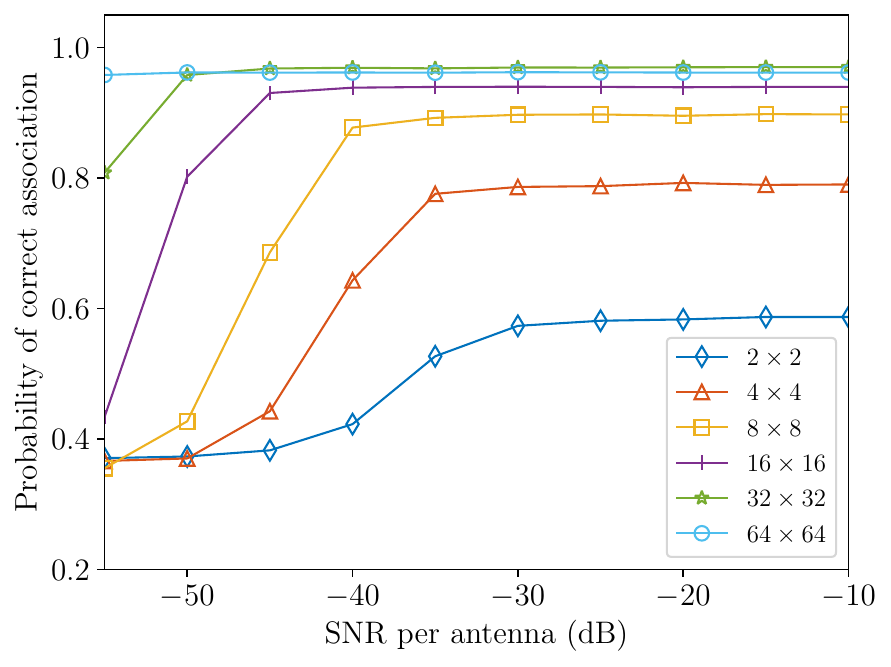}
    \caption{Probability of correct association for varying SNR per antenna utilized to determine the horizontal and vertical analog precoding beam indices selected at the BS. Performance is averaged over the three considered scenarios.}
    \label{fig:probability_of_correct_association}
\end{figure}

We assess in Fig. \ref{fig:probability_of_correct_association} the performance of the proposed T2U association method. The association performance is provided in terms of the probability of correct association between radar targets and VEs averaged over the three considered urban scenarios, as described in Section \ref{sec:t2u_association}. We evaluate the association procedure over increasing SNR per antenna values and we compare the performance on different antenna array dimensions at the BS. Probability of correct association converges to a plateau with the increase of the SNR per antenna. The SNR per antenna can highly influence the beamspace patterns that enable T2U association.  Indeed, faster convergence to the plateau is achieved for larger antenna arrays at the BS, which is expected owing to the greater beamforming gain counteracting lower SNRs per antenna. The achieved probability of correct association plateau rapidly increases with the number of antennas at the BS. This is in accordance with the capability of larger antenna arrays to more accurately \textit{resolve} different targets in the beamspace. As a result, massive antenna arrays show higher resilience to low SNR conditions owing to their significant beamforming gain while providing notable performance in probability of correct association.

\begin{figure}
    \centering
    \includegraphics[width=0.95\columnwidth]{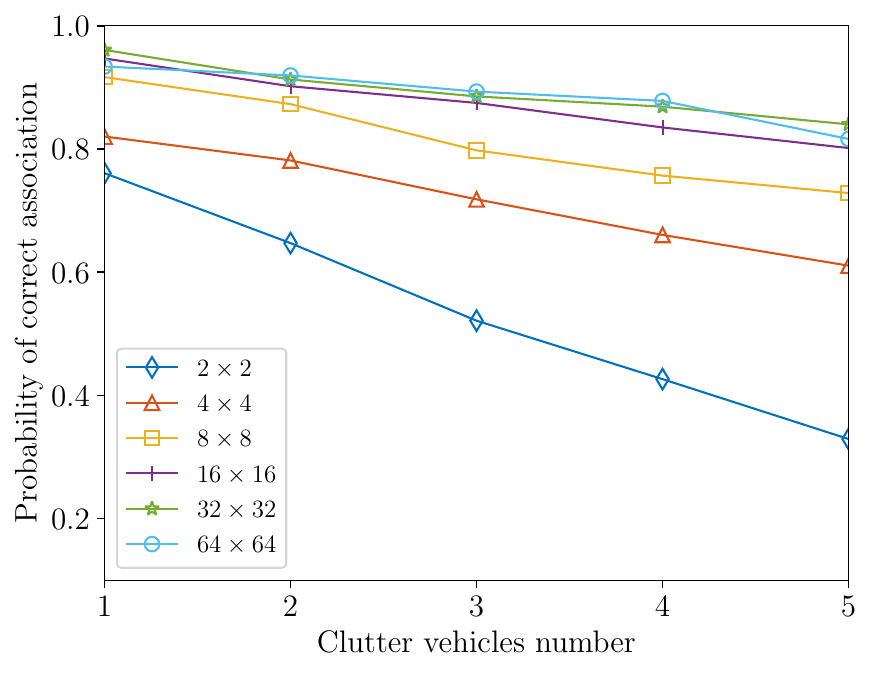}
    \caption{Probability of correct association at -20 dB SNR per antenna for varying number of clutter vehicles and considering association between the detected radar targets and 2 active VEs in the scenario. Performance is averaged over the three considered scenarios.}
    \label{fig:probability_of_correct_association_clutter}
\end{figure}

Fig. \ref{fig:probability_of_correct_association_clutter} studies the variation of the probability of correct association over different dynamic clutter conditions. We consider the association between the detected radar targets and 2 sensed VEs crossing the environment. The remaining vehicles in the scene are treated as dynamic clutter, i.e., (possibly) detected vehicular radar targets which are not active VEs. The association performance shows a decreasing behavior over an increasing number of clutter vehicles in the scene, highlighting the limits of lower dimensional antenna arrays in distinguishing the VEs in the beamspace. Indeed, the probability of association performance significantly improves over increasing sizes of the antenna array utilized at the BS, with an improvement of $\approx48.7\%$ between $2\times2$ and $32\times32$ antenna arrays when 5 clutter vehicles within the sensed environment are considered. Over the simulated vehicular scenarios, no relevant gain in probability of correct association has been observed beyond the $32\times32$ antenna array size.

\begin{figure}
    \centering
    \includegraphics[width=0.95\columnwidth]{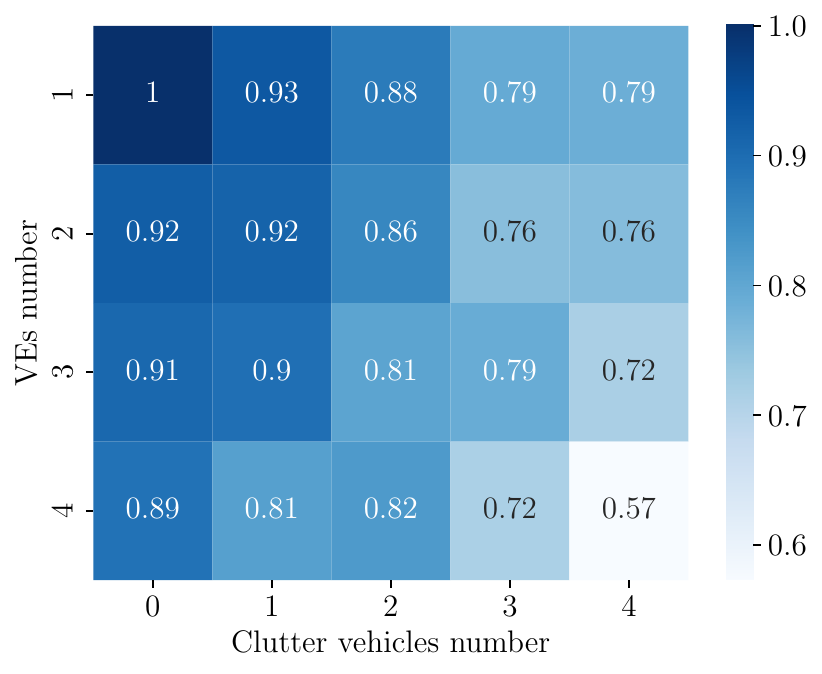}
    \caption{Probability of correct association at -20 dB SNR per antenna for varying number of clutter vehicles considering association between the detected radar targets and a varying number of active VEs (from 1 to 4) in the scenario. The antenna array dimension is set to $8 \times 8$. Performance is averaged over the three considered scenarios.}
    \label{fig:probability_of_correct_association_matrix}
\end{figure}

In Fig. \ref{fig:probability_of_correct_association_matrix}, we provide a matrix showing the probability of correctly associating VEs to vehicular radar targets for a varying number of VEs (from 1 to 4) and of vehicles inducing dynamic clutter in the scene. The matrix is presented for an antenna array of dimensions $8\times 8$ at the BS, and -20 dB SNR per antenna. The first column of the matrix corresponds to the case in which only vehicular targets which are also VEs are crossing the scenario. It represents an upper bound for the following columns. In general, a decreasing behavior in the probability of correct association is observed for an increase in the number of clutter vehicles in the environment. A similar behavior can be observed over the matrix rows for an increasing number of VEs in the scenario. Therefore, as expected, the uncertainty in the matching raises with the number of both VEs and clutter vehicles present in the scene. We observed an equivalent behavior for the remaining antenna array dimensions ($2\times 2$, $4\times 4$, $16\times 16$, $32\times 32$, $64\times 64$) considered in Figs. \ref{fig:probability_of_correct_association} and \ref{fig:probability_of_correct_association_clutter}. In particular, the decreasing trend is more marked for lower dimensional antenna arrays, as envisaged by Fig. \ref{fig:probability_of_correct_association_clutter}. We observed an analogous attitude over different SNRs per antenna between -55 dB and -10 dB. In the latter case, the values in the matrix experience a heavy reduction when approaching lower SNRs per antenna. The same behavior can be observed in Fig. \ref{fig:probability_of_correct_association}.

The derivation of a theoretical expression for the probability of correct association and its theoretical relationship in the beamspace with multi-target detection is a complex task when generic 2D input radar images are considered. We believe that, owing to its complexity, it can pertain to a completely new contribution that we consider for prospective research, complementary to the empirical study here proposed. This investigation of the target to user association problem from a data-driven point of view has been motivated by the unparalleled empirical performance of deep learning models on the extraction of meaningful features from complex structured data. Therefore, we focused on the identification of the architectural biases of deep learning models \cite{goyal2022inductive} faithfully representing the problem at hand. Owing to the pattern recognition capabilities of the latest deep learning models for multi-target detection and classification \cite{xiao2020review}, and to the flexibility provided by the specification of a loss function to use for the optimization of learning models, we have focused on the empirical evaluation of target-to-user association when radar multi-target detection/classification and beam prediction are considered as end-to-end learned downstream tasks. To pursue a fair empirical evaluation of the proposed method, we designed a realistic simulation framework comprising vehicular simulations and dynamic radar and wireless communication simulations, in order to train and test the proposed models across data considering the presence of dynamic complex vehicle meshes in the environment.

\section{Conclusion}\label{sec:conclusion}
In this paper, we addressed target-to-user (T2U) association in next-generation V2I ICAS systems presenting a DL-aided method to associate vehicular radar targets to communication users in the communication beamspace.
We modeled the beam prediction problem as a classification task over a fixed-size codebook of beamforming vectors, and we modified the renowned YOLO architecture to infer both the radar targets' classes and the beamforming vectors' classes for each detected target. 
We trained the YOLO model over recurring vehicular motion patterns---and, therefore, VEs' patterns in the beamspace---within the sensed urban scenario. The aim of the proposed method is to provide the communication system with information on the matching between sensed radar targets and communication users, which is essential to extensively take advantage of the ICAS system capabilities and to enhance the performance of common communication tasks, e.g., beam prediction and blockage prediction.

Simulation results over different vehicular mobility scenarios using realistic mmWave radar and communication ray-tracing data and vehicular traffic show that the proposed method achieves significant T2U association performance, matching the detected radar targets to the respective communication users with a probability of correct association increasing with the antenna array size at the BS---highlighting the respective increase of the separability of the VEs in the beamspace. Moreover, we show that the modified YOLO architecture can perform both beam prediction and radar target detection, with similar performance in mean average precision on the latter over different antenna array sizes. This allows us to effectively predict both target positional information (in terms of bounding boxes in the radar image) and classification, and beam prediction starting from the observation of the whole radar image and, therefore, jointly considering the presence of multiple radar targets in the environment. As a future work, we plan to investigate the integration of radar multi-target tracking by keeping information on the identity of the detected vehicular targets over consecutive time steps, thus reducing the T2U association frequency and increasing the efficiency of the association workflow.

\section*{Acknowledgment}
The research has been carried out in the framework of the Joint Lab between Huawei and Politecnico di Milano.

\bibliographystyle{IEEEtran}
\bibliography{Bibliography}

\begin{thebibliography}{10}
\providecommand{\url}[1]{#1}
\csname url@samestyle\endcsname
\providecommand{\newblock}{\relax}
\providecommand{\bibinfo}[2]{#2}
\providecommand{\BIBentrySTDinterwordspacing}{\spaceskip=0pt\relax}
\providecommand{\BIBentryALTinterwordstretchfactor}{4}
\providecommand{\BIBentryALTinterwordspacing}{\spaceskip=\fontdimen2\font plus
\BIBentryALTinterwordstretchfactor\fontdimen3\font minus \fontdimen4\font\relax}
\providecommand{\BIBforeignlanguage}[2]{{%
\expandafter\ifx\csname l@#1\endcsname\relax
\typeout{** WARNING: IEEEtran.bst: No hyphenation pattern has been}%
\typeout{** loaded for the language `#1'. Using the pattern for}%
\typeout{** the default language instead.}%
\else
\language=\csname l@#1\endcsname
\fi
#2}}
\providecommand{\BIBdecl}{\relax}
\BIBdecl

\bibitem{liu2022integrated}
F.~Liu, Y.~Cui, C.~Masouros, J.~Xu, T.~X. Han, Y.~C. Eldar, and S.~Buzzi, ``{Integrated Sensing and Communications: Toward Dual-Functional Wireless Networks for 6G and Beyond},'' \emph{IEEE Journal on Selected Areas in Communications}, 2022.

\bibitem{huang2021mimo}
S.~Huang, M.~Zhang, Y.~Gao, and Z.~Feng, ``{MIMO Radar Aided mmWave Time-Varying Channel Estimation in MU-MIMO V2X Communications},'' \emph{IEEE Transactions on Wireless Communications}, vol.~20, no.~11, pp. 7581--7594, 2021.

\bibitem{demirhan2022radar}
U.~Demirhan and A.~Alkhateeb, ``{Radar Aided 6G Beam Prediction: Deep Learning Algorithms and Real-World Demonstration},'' in \emph{2022 IEEE Wireless Communications and Networking Conference (WCNC)}.\hskip 1em plus 0.5em minus 0.4em\relax IEEE, 2022, pp. 2655--2660.

\bibitem{gonzalez2016radar}
N.~Gonzalez-Prelcic, R.~M{\'e}ndez-Rial, and R.~W. Heath, ``{Radar Aided Beam Alignment in MmWave V2I Communications Supporting Antenna Diversity},'' in \emph{2016 Information Theory and Applications Workshop (ITA)}.\hskip 1em plus 0.5em minus 0.4em\relax IEEE, 2016, pp. 1--7.

\bibitem{zhang2021overview}
J.~A. Zhang, F.~Liu, C.~Masouros, R.~W. Heath, Z.~Feng, L.~Zheng, and A.~Petropulu, ``{An Overview of Signal Processing Techniques for Joint Communication and Radar Sensing},'' \emph{IEEE Journal of Selected Topics in Signal Processing}, vol.~15, no.~6, pp. 1295--1315, 2021.

\bibitem{9502647}
Y.~Heng, J.~G. Andrews, J.~Mo, V.~Va, A.~Ali, B.~L. Ng, and J.~C. Zhang, ``{Six Key Challenges for Beam Management in 5.5G and 6G Systems},'' \emph{IEEE Communications Magazine}, vol.~59, no.~7, pp. 74--79, 2021.

\bibitem{mizmizi2022fastening}
M.~Mizmizi, F.~Linsalata, M.~Brambilla, F.~Morandi, K.~Dong, M.~Magarini, M.~Nicoli, M.~N. Khormuji, P.~Wang, R.~A. Pitaval \emph{et~al.}, ``{Fastening the initial access in 5G NR sidelink for 6G V2X networks},'' \emph{Vehicular Communications}, vol.~33, p. 100402, 2022.

\bibitem{9838711}
K.~Dong, M.~Mizmizi, D.~Tagliaferri, and U.~Spagnolini, ``{Vehicular Blockage Modelling and Performance Analysis for mmWave V2V Communications},'' in \emph{ICC 2022 - IEEE International Conference on Communications}, 2022, pp. 3604--3609.

\bibitem{ganji2022beamsurfer}
V.~S.~S. Ganji, T.-H. Lin, F.~A. Espinal, and P.~Kumar, ``{BeamSurfer: Minimalist Beam Management of Mobile mm-Wave Devices},'' \emph{IEEE Transactions on Wireless Communications}, vol.~21, no.~11, pp. 8935--8949, 2022.

\bibitem{morandi2021probabilistic}
F.~Morandi, F.~Linsalata, M.~Brambilla, M.~Mizmizi, M.~Magarini, and U.~Spagnolini, ``{A Probabilistic Codebook Technique for Fast Initial Access in 6G Vehicle-to-Vehicle Communications},'' in \emph{2021 IEEE International Conference on Communications Workshops (ICC Workshops)}.\hskip 1em plus 0.5em minus 0.4em\relax IEEE, 2021, pp. 1--6.

\bibitem{alrabeiah2020deep}
M.~Alrabeiah and A.~Alkhateeb, ``{Deep Learning for mmWave Beam and Blockage Prediction Using Sub-6 GHz Channels},'' \emph{IEEE Transactions on Communications}, vol.~68, no.~9, pp. 5504--5518, 2020.

\bibitem{khan2020position}
M.~S. Khan, Q.~Sultan, and Y.~S. Cho, ``{Position and Machine Learning-Aided Beam Prediction and Selection Technique in Millimeter-Wave Cellular System},'' in \emph{2020 International Conference on Information and Communication Technology Convergence (ICTC)}.\hskip 1em plus 0.5em minus 0.4em\relax IEEE, 2020, pp. 603--605.

\bibitem{girshick2014rich}
R.~Girshick, J.~Donahue, T.~Darrell, and J.~Malik, ``{Rich Feature Hierarchies for Accurate Object Detection and Semantic Segmentation},'' in \emph{Proceedings of the IEEE conference on computer vision and pattern recognition}, 2014, pp. 580--587.

\bibitem{liu2016ssd}
W.~Liu, D.~Anguelov, D.~Erhan, C.~Szegedy, S.~Reed, C.-Y. Fu, and A.~C. Berg, ``Ssd: Single shot multibox detector,'' in \emph{Computer Vision--ECCV 2016: 14th European Conference, Amsterdam, The Netherlands, October 11--14, 2016, Proceedings, Part I 14}.\hskip 1em plus 0.5em minus 0.4em\relax Springer, 2016, pp. 21--37.

\bibitem{redmon2016you}
J.~Redmon, S.~Divvala, R.~Girshick, and A.~Farhadi, ``{You Only Look Once: Unified, Real-Time Object Detection},'' in \emph{Proceedings of the IEEE Conference on Computer Vision and Pattern Recognition}, 2016, pp. 779--788.

\bibitem{he2015spatial}
K.~He, X.~Zhang, S.~Ren, and J.~Sun, ``{Spatial Pyramid Pooling in Deep Convolutional Networks for Visual Recognition},'' \emph{IEEE Transactions on Pattern Analysis and Machine Intelligence}, vol.~37, no.~9, pp. 1904--1916, 2015.

\bibitem{girshick2015fast}
R.~Girshick, ``{Fast R-CNN},'' in \emph{Proceedings of the IEEE International Conference on Computer Vision}, 2015, pp. 1440--1448.

\bibitem{ren2015faster}
S.~Ren, K.~He, R.~Girshick, and J.~Sun, ``{Faster R-CNN: Towards Real-Time Object Detection with Region Proposal Networks},'' \emph{Advances in Neural Information Processing Systems}, vol.~28, 2015.

\bibitem{lin2017feature}
T.-Y. Lin, P.~Doll{\'a}r, R.~Girshick, K.~He, B.~Hariharan, and S.~Belongie, ``{Feature Pyramid Networks for Object Detection},'' in \emph{Proceedings of the IEEE Conference on Computer Vision and Pattern Recognition}, 2017, pp. 2117--2125.

\bibitem{yolov8}
``{Ultralytics YOLOv8},'' \url{ https://github.com/ultralytics/ultralytics }, {Accessed on April 2023}.

\bibitem{zou2023object}
Z.~Zou, K.~Chen, Z.~Shi, Y.~Guo, and J.~Ye, ``{Object Detection in 20 Years: A Survey},'' \emph{Proceedings of the IEEE}, 2023.

\bibitem{kim2020yolo}
W.~Kim, H.~Cho, J.~Kim, B.~Kim, and S.~Lee, ``{YOLO-Based Simultaneous Target Detection and Classification in Automotive FMCW Radar Systems},'' \emph{Sensors}, vol.~20, no.~10, p. 2897, 2020.

\bibitem{long2020lira}
Z.~Long, W.~Suyuan, C.~Zhongma, F.~Jiaqi, Y.~Xiaoting, and D.~Wei, ``Lira-yolo: a lightweight model for ship detection in radar images,'' \emph{Journal of Systems Engineering and Electronics}, vol.~31, no.~5, pp. 950--956, 2020.

\bibitem{song2022ms}
Y.~Song, Z.~Xie, X.~Wang, and Y.~Zou, ``{MS-YOLO: Object Detection Based on YOLOv5 Optimized Fusion Millimeter-Wave Radar and Machine Vision},'' \emph{IEEE Sensors Journal}, vol.~22, no.~15, pp. 15\,435--15\,447, 2022.

\bibitem{mizmizi2023target}
M.~Mizmizi, D.~Tagliaferri, D.~Badini, and U.~Spagnolini, ``{Target-to-User Association in ISAC Systems With Vehicle-Lodged RIS},'' \emph{IEEE Wireless Communications Letters}, 2023.

\bibitem{aydogdu2020distributed}
C.~Aydogdu, F.~Liu, C.~Masouros, H.~Wymeersch, and M.~Rydstr{\"o}m, ``{Distributed Radar-aided Vehicle-to-Vehicle Communication},'' in \emph{2020 IEEE Radar Conference (RadarConf20)}.\hskip 1em plus 0.5em minus 0.4em\relax IEEE, 2020, pp. 1--6.

\bibitem{wang2022multi}
Z.~Wang, K.~Han, J.~Jiang, F.~Liu, and W.~Yuan, ``{Multi-Vehicle Tracking and ID Association Based on Integrated Sensing and Communication Signaling},'' \emph{IEEE Wireless Communications Letters}, vol.~11, no.~9, pp. 1960--1964, 2022.

\bibitem{sumo}
``{Simulation of Urban Mobility},'' \url{ https://www.eclipse.org/sumo/ }, {Accessed on May 2023}.

\bibitem{wireless_insite}
``{Remcom Wireless InSite},'' \url{ https://www.remcom.com/wireless-insite-em-propagation-software }, {Accessed on May 2023}.

\bibitem{wavefarer}
``{Remcom WaveFarer},'' \url{ https://www.remcom.com/wavefarer-automotive-radar-software }, {Accessed on May 2023}.

\bibitem{6717211}
O.~E. Ayach, S.~Rajagopal, S.~Abu-Surra, Z.~Pi, and R.~W. Heath, ``Spatially sparse precoding in millimeter wave mimo systems,'' \emph{IEEE Transactions on Wireless Communications}, vol.~13, no.~3, pp. 1499--1513, 2014.

\bibitem{6834753}
M.~R. {Akdeniz}, Y.~{Liu}, M.~K. {Samimi}, S.~{Sun}, S.~{Rangan}, T.~S. {Rappaport}, and E.~{Erkip}, ``Millimeter wave channel modeling and cellular capacity evaluation,'' \emph{IEEE Journal on Selected Areas in Communications}, vol.~32, no.~6, pp. 1164--1179, 2014.

\bibitem{mizmizi2021channel}
M.~Mizmizi, D.~Tagliaferri, D.~Badini, C.~Mazzucco, and U.~Spagnolini, ``{Channel Estimation for 6G V2X Hybrid Systems Using Multi-Vehicular Learning},'' \emph{IEEE Access}, vol.~9, pp. 95\,775--95\,790, 2021.

\bibitem{Zaugg2015_FMCWSAR}
E.~C. {Zaugg} and D.~G. {Long}, ``{Generalized Frequency Scaling and Backprojection for LFM-CW SAR Processing},'' \emph{IEEE Transactions on Geoscience and Remote Sensing}, vol.~53, no.~7, pp. 3600--3614, 2015.

\bibitem{skolnik}
M.~Skolnik, \emph{Introduction to Radar Systems}.\hskip 1em plus 0.5em minus 0.4em\relax London: McGraw-Hill Education, 2002.

\bibitem{Cafforio1991}
C.~{Cafforio}, C.~{Prati}, and F.~{Rocca}, ``Sar data focusing using seismic migration techniques,'' \emph{IEEE Transactions on Aerospace and Electronic Systems}, vol.~27, no.~2, pp. 194--207, 1991.

\bibitem{5739256}
J.~Gunther, R.~West, N.~Crookston, and T.~Moon, ``Maximum likelihood synthetic aperture radar image formation for highly nonlinear flight tracks,'' in \emph{2011 Digital Signal Processing and Signal Processing Education Meeting (DSP/SPE)}, 2011, pp. 449--454.

\bibitem{wang2020cspnet}
C.-Y. Wang, H.-Y.~M. Liao, Y.-H. Wu, P.-Y. Chen, J.-W. Hsieh, and I.-H. Yeh, ``{CSPNet: A New Backbone That Can Enhance Learning Capability of CNN},'' in \emph{Proceedings of the IEEE/CVF conference on computer vision and pattern recognition workshops}, 2020, pp. 390--391.

\bibitem{wang2023yolov7}
C.-Y. Wang, A.~Bochkovskiy, and H.-Y.~M. Liao, ``{YOLOv7: Trainable Bag-of-Freebies Sets New State-of-the-Art for Real-Time Object Detectors},'' in \emph{Proceedings of the IEEE/CVF Conference on Computer Vision and Pattern Recognition}, 2023, pp. 7464--7475.

\bibitem{liu2018path}
S.~Liu, L.~Qi, H.~Qin, J.~Shi, and J.~Jia, ``{Path Aggregation Network for Instance Segmentation},'' in \emph{Proceedings of the IEEE Conference on Computer Vision and Pattern Recognition}, 2018, pp. 8759--8768.

\bibitem{feng2021tood}
C.~Feng, Y.~Zhong, Y.~Gao, M.~R. Scott, and W.~Huang, ``{TOOD: Task-aligned One-stage Object Detection},'' in \emph{2021 IEEE/CVF International Conference on Computer Vision (ICCV)}.\hskip 1em plus 0.5em minus 0.4em\relax IEEE Computer Society, 2021, pp. 3490--3499.

\bibitem{li2023generalized}
X.~Li, C.~Lv, W.~Wang, G.~Li, L.~Yang, and J.~Yang, ``{Generalized Focal Loss: Towards Efficient Representation Learning for Dense Object Detection},'' \emph{IEEE Transactions on Pattern Analysis and Machine Intelligence}, vol.~45, no.~3, pp. 3139--3153, 2023.

\bibitem{zheng2020distance}
Z.~Zheng, P.~Wang, W.~Liu, J.~Li, R.~Ye, and D.~Ren, ``{Distance-IoU Loss: Faster and Better Learning for Bounding Box Regression},'' in \emph{Proceedings of the AAAI Conference on Artificial Intelligence}, vol.~34, no.~07, 2020, pp. 12\,993--13\,000.

\bibitem{openstreetmap}
``{OpenStreetMap},'' \url{ https://www.openstreetmap.org/ }, {Accessed on May 2023}.

\bibitem{blender}
``{Blender},'' \url{ https://www.blender.org/ }, {Accessed on May 2023}.

\bibitem{blenderosm}
``{Blender-OSM},'' \url{ https://prochitecture.gumroad.com/l/blender-osm }, {Accessed on May 2023}.

\bibitem{degli2021ray}
V.~Degli~Esposti, ``{Ray tracing: techniques, applications and prospect},'' in \emph{2020 International Symposium on Antennas and Propagation (ISAP)}.\hskip 1em plus 0.5em minus 0.4em\relax IEEE, 2021, pp. 307--308.

\bibitem{he2016deep}
K.~He, X.~Zhang, S.~Ren, and J.~Sun, ``Deep residual learning for image recognition,'' in \emph{Proceedings of the IEEE conference on computer vision and pattern recognition}, 2016, pp. 770--778.

\bibitem{simonyan2014very}
K.~Simonyan and A.~Zisserman, ``Very deep convolutional networks for large-scale image recognition,'' in \emph{3rd International Conference on Learning Representations (ICLR 2015)}.\hskip 1em plus 0.5em minus 0.4em\relax Computational and Biological Learning Society, 2015.

\bibitem{mmdetection}
K.~Chen, J.~Wang, J.~Pang, Y.~Cao, Y.~Xiong, X.~Li, S.~Sun, W.~Feng, Z.~Liu, J.~Xu, Z.~Zhang, D.~Cheng, C.~Zhu, T.~Cheng, Q.~Zhao, B.~Li, X.~Lu, R.~Zhu, Y.~Wu, J.~Dai, J.~Wang, J.~Shi, W.~Ouyang, C.~C. Loy, and D.~Lin, ``{MMDetection}: Open mmlab detection toolbox and benchmark,'' \emph{arXiv preprint arXiv:1906.07155}, 2019.

\bibitem{stroescu2021object}
A.~Stroescu, L.~Daniel, D.~Phippen, M.~Cherniakov, and M.~Gashinova, ``Object detection on radar imagery for autonomous driving using deep neural networks,'' in \emph{2020 17th European Radar Conference (EuRAD)}.\hskip 1em plus 0.5em minus 0.4em\relax IEEE, 2021, pp. 120--123.

\bibitem{goyal2022inductive}
A.~Goyal and Y.~Bengio, ``Inductive biases for deep learning of higher-level cognition,'' \emph{Proceedings of the Royal Society A}, vol. 478, no. 2266, p. 20210068, 2022.

\bibitem{xiao2020review}
Y.~Xiao, Z.~Tian, J.~Yu, Y.~Zhang, S.~Liu, S.~Du, and X.~Lan, ``A review of object detection based on deep learning,'' \emph{Multimedia Tools and Applications}, vol.~79, pp. 23\,729--23\,791, 2020.

\end{thebibliography}

\end{document}